\documentclass[twocolumn,aps,prl,floatfix,amsmath,amssymb,superscriptaddress,longbibliography,nofootinbib]{revtex4-2}
\usepackage{graphicx}
\usepackage[dvipsnames]{xcolor}
\usepackage{amsthm}
\definecolor{PRLBlue}{RGB}{46,48,146}
\usepackage[
    colorlinks=true,
    citecolor=PRLBlue,
    linkcolor=PRLBlue,
    urlcolor=PRLBlue,
]{hyperref}

\newcommand{\bra}[1]{\left\langle#1\right|}
\newcommand{\ket}[1]{\left|#1\right\rangle}
\allowdisplaybreaks

\numberwithin{equation}{section}

\theoremstyle{plain}

\begin{document}

\title{\texorpdfstring{Energy-resolved current response and the crossover between two approximately linear-in-$T$ resistivity regimes in the infinite-$U$ Hubbard model}{Energy-resolved current response and the crossover between two approximately linear-in-T resistivity regimes in the infinite-U Hubbard model}}

\author{Shuo Liu}
\affiliation{Department of Physics, Princeton University, Princeton, New Jersey 08544, USA}

\author{Aman Kumar}
\affiliation{National High Magnetic Field Laboratory, Tallahassee, Florida 32310, USA}
\affiliation{Department of Physics, Florida State University, Tallahassee, Florida 32306, USA}
\affiliation{FSU Quantum Initiative, Florida State University, 282 Champions Way, Tallahassee, FL 32306, United States}
\affiliation{Division of Physics and Applied Physics, School of Physical and Mathematical Sciences,
Nanyang Technological University, Singapore 637371}
\author{Yuhao Ma}
\affiliation{Department of Physics and Institute of Condensed Matter Theory, University of Illinois at Urbana-Champaign, Urbana, IL 61801, USA}

\author{Philip W. Phillips}
\affiliation{Department of Physics and Institute of Condensed Matter Theory, University of Illinois at Urbana-Champaign, Urbana, IL 61801, USA}

\author{Hitesh J. Changlani}
\affiliation{National High Magnetic Field Laboratory, Tallahassee, Florida 32310, USA}
\affiliation{Department of Physics, Florida State University, Tallahassee, Florida 32306, USA}
\affiliation{FSU Quantum Initiative, Florida State University, 282 Champions Way, Tallahassee, FL 32306, United States}

\author{B. Andrei Bernevig}
\email{bernevig@princeton.edu}
\affiliation{Department of Physics, Princeton University, Princeton, New Jersey 08544, USA}
\affiliation{Donostia International Physics Center, P. Manuel de Lardizabal 4, 20018 Donostia-San Sebastian, Spain}
\affiliation{IKERBASQUE, Basque Foundation for Science, Bilbao, Spain}
\affiliation{Center for Computational Quantum Physics, Flatiron Institute, New York, New York 10010, USA}

\date{\today}

\begin{abstract}
We study charge transport in the two-dimensional infinite-$U$ Hubbard model with hopping amplitude $t$ using full exact diagonalization (ED) on $N=16$- and $18$-site square clusters with a single doped hole. We first provide a proof, treating degeneracies explicitly, that a constant dc $f_x$ function, defined as the sum of the squared current matrix elements connecting eigenstates at a given energy to eigenstates of nearly the same energy, averaged over degenerate initial states, is necessary and sufficient for strictly linear-in-$T$ resistivity: $\rho_x(T)=\alpha T$ for all $T>0$ with a constant $\alpha$. The numerically extrapolated dc $f_x$ is not exactly constant, but it is approximately flat over a broad region in the interior of the many-body spectrum and varies strongly within a scale $\varepsilon_{f_x}$ of the spectral edges. This scale $\varepsilon_{f_x}$ is related to a characteristic temperature $T_{\mathrm{cr}}$, which quantifies the leading departure of the resistivity from high-temperature linearity upon cooling. Within the studied temperature range $T/t \geq 0.05$, our ED resistivity results reveal a gradual crossover between high- and low-temperature approximately linear-in-$T$ regimes, with characteristic crossover temperatures $T_{\mathrm{cr}}/t\approx0.414$ and $0.467$ for $N=16$ and $18$, respectively. We also find a broad specific-heat peak associated with this gradual crossover, at $T/t \approx 0.70$ and $0.68$ for $N=16$ and $18$, respectively.
\end{abstract}

\maketitle

\renewcommand{\theequation}{\arabic{equation}}

Strange metals, characterized in part by a resistivity that is
approximately linear in temperature over an extended range,
pose a central puzzle in correlated-electron
physics~\cite{Phillips2022}. Such dc transport behavior has been
observed in cuprate superconductors~\cite{PhysRevLett.81.4720,PhysRevLett.92.167001,jinLinkSpinFluctuations2011,PhysRevB.41.846,doi:10.1126/science.1165015,daouLinearTemperatureDependence2009,PhysRevB.95.224517,doiron-leyraudPseudogapPhaseCuprate2017,legrosUniversalTlinearResistivity2019,michonReconcilingScalingOptical2023},
infinite-layer nickelates~\cite{leeLinearintemperatureResistivityOptimally2023,hsuTransportPhaseDiagram2024},
and twisted bilayer graphene~\cite{polshynLargeLinearintemperatureResistivity2019,PhysRevLett.124.076801,jaouiQuantumCriticalBehaviour2022}, albeit with a nonzero intercept at zero temperature.
Its occurrence across a wide variety of materials and cold-atom platforms suggests a possible
common mechanism, while its microscopic
origin and a minimal physical model that captures
it remain subjects of active research.

The two-dimensional (2D) Hubbard
model~\cite{doi:10.1126/science.235.4793.1196,PhysRevB.37.3759,annurev:/content/journals/10.1146/annurev-conmatphys-090921-033948,annurev:/content/journals/10.1146/annurev-conmatphys-031620-102024}
is widely regarded as a minimal model for cuprate
superconductors~\cite{RevModPhys.66.763,RevModPhys.75.473,RevModPhys.82.1719,doi:10.1073/pnas.1316512110,RevModPhys.87.457,keimer_quantum_2015,PhysRevB.39.6880}. With the development of numerical methods such as quantum Monte Carlo~\cite{PhysRevB.41.9301,PhysRevLett.94.170201,PhysRevB.71.155115,PhysRevB.91.241117,PhysRevLett.117.267002,PhysRevLett.116.250601,annurev:/content/journals/10.1146/annurev-conmatphys-033117-054307,PhysRevB.106.L241109,PhysRevB.101.045108} and neural-network approaches~\cite{roth2025superconductivitytwodimensionalhubbardmodel,chen2025neuralnetworkaugmentedpfaffianwavefunctions,rende2026superconductivitytthubbardmodel, viteritti2026variationalbiasresolvingintertwined, gu2025solvinghubbardmodelneural}, the ground state and finite-temperature phase transitions~\cite{PhysRevLett.110.216405,roth2025superconductivitytwodimensionalhubbardmodel,chen2025neuralnetworkaugmentedpfaffianwavefunctions, rende2026superconductivitytthubbardmodel, PhysRevB.47.7995,PhysRevX.11.031007,PhysRevB.82.155101,PhysRevX.8.021048,doi:10.1126/science.ade9194, PhysRevX.13.011007,sherif2025haertershastrykineticmagnetismmetallicity,PhysRevB.79.045133, lu2026thermalsignaturesslatermottcrossover,song2026computationthermalentropydoped,  y7fg-lhsm, PhysRevB.75.085108}, as well as charge transport~\cite{PhysRevB.106.085142, PhysRevB.42.10807, PhysRevB.80.161105,PhysRevLett.123.036601,   PhysRevB.102.115142,PhysRevB.104.165143,PhysRevB.95.041110,PhysRevLett.110.086401,doi:10.1126/science.aau7063,fratini2026strangemetaltransportcoupling}, have been studied in this strongly correlated system. Hubbard physics has also been studied in other dimensions~\cite{PhysRevB.8.312, PhysRevB.51.10386, PhysRevLett.20.1445, PhysRevB.79.035112,essler2005one, PhysRevLett.56.1529, gohmann1998fermionic, 10.1143/PTP.52.103, IZERGIN1998537, PhysRevB.41.2326, liu2026transporttemperature1exact,PhysRevB.111.035123,PhysRevLett.134.016503,PhysRevB.48.7167,PhysRevB.45.6479, PhysRevB.94.235115, PhysRevB.96.054114, PhysRevLett.69.1240,RevModPhys.68.13}, as well as in related models~\cite{10.21468/SciPostPhys.6.6.067,sharma2025instabilitynagaokastatequantum,PhysRevX.14.041025,PhysRevLett.68.2402,PhysRevLett.64.475, PhysRevB.55.6504,PhysRevB.52.6903,PhysRevB.59.5341}, such as
moir\'e Hubbard models~\cite{PhysRevResearch.2.033087, PhysRevLett.121.026402, wangCorrelatedElectronicPhases2020, tuo2026chiralspinliquidchiral, PhysRevX.12.021064}.
Besides the linear-in-$T$ resistivity in high-temperature regimes observed numerically~\cite{PhysRevB.102.115142,PhysRevB.104.165143,PhysRevB.95.041110,PhysRevLett.110.086401} and in cold-atom experiments~\cite{doi:10.1126/science.aat4134}, approximately linear-in-$T$ resistivity over broad
temperature ranges relevant to strange-metal
phenomenology has also been observed in the 2D Hubbard model with finite interaction $U$~\cite{doi:10.1126/science.aau7063} and in the infinite-$U$ limit~\cite{fratini2026strangemetaltransportcoupling}. The infinite-$U$ limit is equivalent to the $t$-$J$ model at $J=0$, and a finite $J$ was found to suppress the linear-in-$T$ resistivity at low temperatures~\cite{fratini2026strangemetaltransportcoupling}, but a many-body explanation of this linear-in-$T$ behavior is still lacking. We note that linear-in-$T$ resistivity also arises in nonlocal
SYK-type models~\cite{RevModPhys.94.035004,doi:10.1126/science.abq6011,doi:10.1073/pnas.2003179117,PhysRevLett.133.186502}. However, what allows approximately
linear-in-$T$ resistivity to persist down to lower temperatures, and what
microscopic scale controls its departure from the high-temperature linear behavior, remain open questions.

In this Letter, we use exact diagonalization (ED) on square clusters~\cite{SciPostPhys.2.1.003,SciPostPhys.7.2.020} to study charge transport in the 2D Hubbard model in the infinite-$U$ limit~\cite{PhysRev.147.392,PhysRevLett.108.126406}. Several exact results exist for this model in various geometries~\cite{PhysRevB.98.180101,glittum_resonant_2025, CASPERS19891033, PhysRevB.107.L140401,  PhysRevResearch.6.013307}, including the Nagaoka theorem~\cite{PhysRev.147.392} and recently identified exact eigenstates~\cite{liu2026transporttemperature1exact,LadderBound}. We consider a single hole doped into clusters of $N=16$ and $18$ sites, corresponding to a carrier density of $1/N$. As shown in 1D~\cite{liu2026transporttemperature1exact}, a fixed number of holes and a finite hole density lead to different behavior: the leading low-temperature correction to the Drude weight is linear in $T$ only for a fixed number of holes. However, due to the lack of exact solutions in 2D, we are limited to numerics.

We build on the energy-resolved current-response function
$f_x$~\cite{PhysRevB.105.L201108} [Eq.~\eqref{eq:f_function}] and prove that, for any given Hamiltonian, a constant dc $f_x$ is a necessary and sufficient condition for strictly linear-in-$T$ resistivity. Our proof, an alternative to the derivation in Ref.~\cite{PhysRevB.105.L201108}, is based on a Vandermonde matrix and treats degeneracies explicitly (Sec.~SI of Supplemental Material (SM)~\cite{SupplementalMaterials}). We then show that, in the infinite-$U$ Hubbard model [Eq.~\eqref{eq:main_H}],
the numerically extrapolated dc $f_x$ is nearly flat in the middle of the many-body spectrum but changes rapidly near its edges, over an energy scale $\varepsilon_{f_x} \approx 0.99$ and $1.11$ for $N=16$ and $18$, respectively. This scale is related to a characteristic temperature $T_{\mathrm{cr}}$, which sets the leading relative correction, $(T_{\mathrm{cr}}/T)^2$, to the high-temperature linear resistivity [Eq.~\eqref{eq:main_highT_rhox_a} and Eq.~\eqref{eq:main_Tcr}].

Our ED results show a gradual crossover from the expected high-temperature linear-in-$T$ behavior~\cite{PhysRevB.102.115142,PhysRevB.104.165143,PhysRevB.95.041110,PhysRevLett.110.086401,PhysRevB.81.054512,PhysRevB.94.235115} to a second approximately linear-in-$T$ regime extending down to $T/t=0.05$. We use $T_{\mathrm{cr}}$ as the characteristic crossover temperature, with $T_{\mathrm{cr}}/t\approx0.414$ and $0.467$ for $N=16$ and $18$, respectively. Here $T/t=0.05$ is the lowest temperature studied for charge transport; at lower temperatures, finite ferromagnetic correlations develop~\cite{fratini2026strangemetaltransportcoupling}, although the infinite system is paramagnetic at any $T>0$~\cite{PhysRevB.43.8779}.
We also calculate the specific heat and find a broad peak associated with this gradual crossover.

\textit{Model and Kubo formula.---}We consider the infinite-$U$ Hubbard model~\cite{PhysRev.147.392,PhysRevLett.108.126406} on a square cluster of $N$ sites:
\begin{equation}
\hat{H}(\boldsymbol{\theta})
=
-t\sum_{\mathbf{r},\sigma}
\sum_{\hat{e}=\hat{x},\hat{y}}
\left(
e^{iA_{\hat{e}}}
\tilde{c}^{\dagger}_{\mathbf{r},\sigma}
\tilde{c}_{\mathbf{r}+\hat{e},\sigma}
+\mathrm{H.c.}
\right),
\label{eq:main_H}
\end{equation}
where
$\tilde{c}_{\mathbf{r},\sigma}
=(1-n_{\mathbf{r},\bar{\sigma}})c_{\mathbf{r},\sigma}$
annihilates a fermion with spin $\sigma$ at site $\mathbf{r}$,
$n_{\mathbf{r},\bar{\sigma}}
=c^{\dagger}_{\mathbf{r},\bar{\sigma}}
c_{\mathbf{r},\bar{\sigma}}$
is the density of the opposite spin $\bar{\sigma}$, and the projection
excludes double occupancy. To reduce finite-size effects, we impose twisted boundary conditions
(TBCs)~\cite{PhysRevB.44.9562,PhysRevB.72.075108,PhysRevLett.65.243}
in Eq.~\eqref{eq:main_H} through a uniform vector potential
$\mathbf{A}_{\boldsymbol{\theta}}=(A_x,A_y)$ determined by the twist angles
$\boldsymbol{\theta}=(\theta_1,\theta_2)$ (Sec.~SI of SM~\cite{SupplementalMaterials}). The Peierls phase along a bond in
the $\hat{e}$ direction is
$A_{\hat{e}}\equiv\mathbf{A}_{\boldsymbol{\theta}}\cdot\hat{e}$, and
$\boldsymbol{\theta}=\mathbf{0}$ corresponds to periodic boundary conditions
(PBCs). We set the lattice spacing to unity and $e=\hbar=k_B=1$, keeping the hopping
amplitude $t$ explicit as the energy scale. We study the conventional
$4\times4$ cluster with $N=16$ and a rotated square cluster with $N=18$ (Sec.~SI of SM~\cite{SupplementalMaterials}), each containing a single doped hole. Under PBCs, the ground state is
the fully polarized Nagaoka ferromagnet with energy
$E_{\mathrm{GS}}=-4t$~\cite{PhysRev.147.392}.

The $x$-direction current operator associated with Eq.~\eqref{eq:main_H} is
\begin{equation}
\hat{J}^{x}(\boldsymbol{\theta})
=
it\sum_{\mathbf{r},\sigma}
\left[
e^{iA_x}
\tilde{c}^{\dagger}_{\mathbf{r},\sigma}
\tilde{c}_{\mathbf{r}+\hat{x},\sigma}
-\mathrm{H.c.}
\right].
\label{eq:current_operator}
\end{equation}
Transport along $x$ and $y$ is equivalent after averaging over a
symmetry-preserving twist grid (Sec.~SI of SM~\cite{SupplementalMaterials}), so we consider
only the $x$ direction. Because the ED spectrum is discrete, we replace the delta
function in the Kubo formula~\cite{kubo1957statistical,PhysRevB.16.2437,PhysRevB.35.8391,PhysRevLett.68.2830,PhysRevB.59.1825,PhysRevLett.134.176501,liu2026transporttemperature1exact}
by a Lorentzian of width $\eta$,
$\delta_\eta(E)=\eta/[\pi(\eta^2+E^2)]$ (Sec.~SI of SM~\cite{SupplementalMaterials}).
For a fixed
$\boldsymbol{\theta}$, let $\ket{n(\boldsymbol{\theta})}$ denote the $n$-th eigenstate of
$\hat{H}(\boldsymbol{\theta})$ with energy $E_n(\boldsymbol{\theta})$, and define the current
matrix elements
$J_{mn}^{x}(\boldsymbol{\theta})
=\bra{m(\boldsymbol{\theta})}\hat{J}^{x}(\boldsymbol{\theta})\ket{n(\boldsymbol{\theta})}$.
The regular finite-frequency part of the optical conductivity is
\begin{align}
\sigma_{\mathrm{reg},\boldsymbol{\theta}}^{xx}(\omega,T,\eta)
&=
\frac{1-e^{-\beta\omega}}{NZ(\boldsymbol{\theta})\omega}
\sum_n e^{-\beta E_n(\boldsymbol{\theta})}
\nonumber\\[-2pt]
&\times
\sum_m
\left|J_{mn}^{x}(\boldsymbol{\theta})\right|^2
\frac{\eta}
{\eta^2+(E_n(\boldsymbol{\theta})+\omega-E_m(\boldsymbol{\theta}))^2},
\label{eq:sigma_regular_with_eta_L_main}
\end{align}
where $\beta=1/T$ and
$Z(\boldsymbol{\theta})=\sum_n e^{-\beta E_n(\boldsymbol{\theta})}$ is the partition function at fixed twist $\boldsymbol{\theta}$. In the
limit $\eta\to0$, Eq.~\eqref{eq:sigma_regular_with_eta_L_main} recovers the
standard delta-function Kubo expression at nonzero frequency~\cite{liu2026transporttemperature1exact} (Sec.~SI of SM~\cite{SupplementalMaterials}). We note that the full optical conductivity also contains a Drude contribution at
zero frequency. For nonintegrable systems, the Drude weight is expected to vanish
in the thermodynamic limit~\cite{PhysRevB.73.035113}, although it can remain finite in finite-size systems (Sec.~SI of SM~\cite{SupplementalMaterials}).

To obtain the dc resistivity under TBCs, we first average
Eq.~\eqref{eq:sigma_regular_with_eta_L_main} over $N_\theta=121$ twists,
$\sigma_{\mathrm{reg}}^{xx}
=N_\theta^{-1}\sum_{\boldsymbol{\theta}}
\sigma_{\mathrm{reg},\boldsymbol{\theta}}^{xx}$
(Sec.~SI of SM~\cite{SupplementalMaterials}), using a fixed $\eta/t=0.005$
throughout (Sec.~SII of SM~\cite{SupplementalMaterials}). We then fit the
twist-averaged conductivity $\sigma_{\mathrm{reg}}^{xx}$ over
$0.025\leq\omega/t\leq0.055$ to
$\sigma_{\mathrm{dc}}^{xx}(T)+c_2(T)(\omega/t)^2$ and take the intercept
$\sigma_{\mathrm{dc}}^{xx}(T)$ as our finite-cluster, finite-$\eta$ estimate
of the dc conductivity. The fit window is restricted to finite frequencies
to mitigate the influence of the spurious finite-size Drude-like peak
(Sec.~SII of SM~\cite{SupplementalMaterials}). The dc resistivity is then
$\rho_x(T)=1/\sigma_{\mathrm{dc}}^{xx}(T)$. 
In principle, the dc resistivity in the thermodynamic
limit is obtained by taking $N\to\infty$ before $\eta\to0$ and $\omega\to0$
(Sec.~SI of SM~\cite{SupplementalMaterials}).

\begin{figure}[t]
\centering
\includegraphics[width=0.95\columnwidth, keepaspectratio]{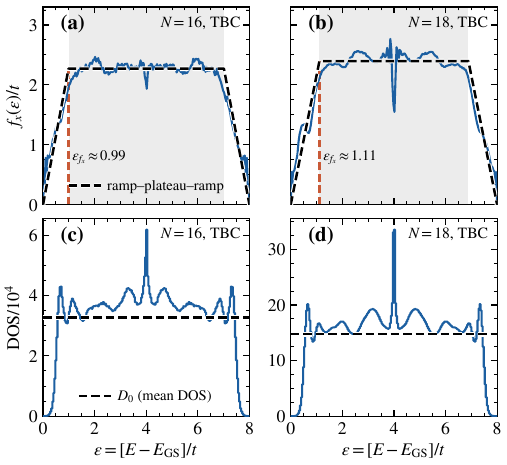}
\caption{Extrapolated dc $f_x(\varepsilon)$ for (a) $N=16$ and (b) $N=18$ under TBCs, as a function of the excitation energy $\varepsilon=(E-E_{\mathrm{GS}})/t$. (c),(d) Density of states (DOS) for $N=16$ and $18$, respectively. Black dashed lines show the approximations used in the simplified model: a ramp--plateau--ramp form in (a),(b) and a constant DOS in (c),(d). Red dashed lines in (a),(b) mark the edge scales $\varepsilon_{f_x}\approx0.99$ and $1.11$, respectively.}
\label{fig:main_fx}
\end{figure}

\begin{figure}[t]
\centering
\includegraphics[width=\columnwidth, keepaspectratio]{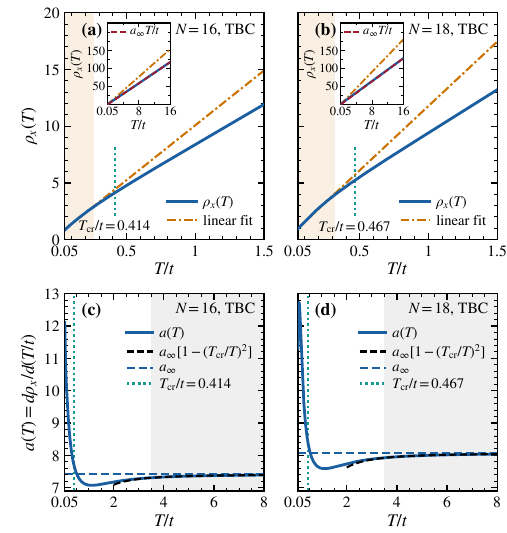}
\caption{(a),(b) dc resistivity $\rho_x(T)$ of the infinite-$U$ Hubbard model with a single doped hole under TBCs for $N=16$ and $18$, respectively. (c),(d) Corresponding local slope $a(T)=d\rho_x/d(T/t)$. Blue dashed lines in (c),(d) mark the numerically obtained high-temperature slope $a_{\infty}$. Black dashed lines in (c),(d) are fits to Eq.~\eqref{eq:main_highT_rhox_a} within $3.5 \leq T/t \leq 8$ (gray shaded region), from which $T_{\mathrm{cr}}$ (green dotted lines in all panels) is extracted. The resistivity shows two approximately linear-in-$T$ regimes connected by a gradual crossover. Orange dash-dotted lines in (a),(b) are linear fits to $\rho_x(T)$ over $0.05 \leq T/t \leq T_{\mathrm{cr}}/t-0.15$ (beige shaded region). Insets in (a),(b) show the full range $0.05 \leq T/t \leq 16$ and compare these linear fits with the high-temperature linear behavior $a_\infty T/t$ (red dashed lines).}
\label{fig:main_rho}
\end{figure}

\textit{General criterion for strictly linear-in-$T$ resistivity.---}Before presenting the ED transport results for Eq.~\eqref{eq:main_H}, we define the energy-resolved
current-response function $f_x$~\cite{PhysRevB.105.L201108,PhysRevB.110.064501} and derive a necessary and sufficient condition on $f_x$ for strictly linear-in-$T$ resistivity (defined below). Since the criterion applies to other models as well, we consider a general Hamiltonian $\hat{H}$ and denote its $n$-th eigenstate by $\ket{n}$, with eigenenergy $E_{n}$. The corresponding $x$-direction current operator is $\hat{J}^{x}$, with matrix elements $J^{x}_{mn} = \bra{m} \hat{J}^{x} \ket{n}$. Let
$\{\varepsilon_i\}_{i=1}^{M}$ denote the $M$ distinct eigenenergies of $\hat{H}$, where $\varepsilon_i$ has degeneracy $g_i\geq1$. We define
\begin{align}
f_x(\varepsilon_i,\omega,\eta)
&=
\frac{1}{g_i}
\sum_{\substack{n\\ E_n=\varepsilon_i}}
\sum_m
\left|J_{mn}^{x}\right|^2
\frac{\eta}
{\eta^2+(E_n+\omega-E_m)^2},
\label{eq:f_function}
\end{align}
where the first sum runs over all degenerate eigenstates with energy $\varepsilon_{i}$. Since $\lvert J_{mn}^{x}\rvert^2$ is the current-transition strength between eigenstates $\ket{n}$ and $\ket{m}$, and the Lorentzian kernel
selects transitions with $E_m-E_n\simeq\omega$ within a resolution $\eta$, $f_x(\varepsilon_i,\omega,\eta)$ measures the current response at energy $\varepsilon_i$ and frequency $\omega$, averaged over the degenerate states.

Combining Eqs.~\eqref{eq:sigma_regular_with_eta_L_main} and~\eqref{eq:f_function}, written for the general Hamiltonian $\hat{H}$ without the twist label $\boldsymbol{\theta}$, we obtain
\begin{align}
\sigma_{\mathrm{reg}}^{xx}(\omega,T,\eta)
&=
\frac{1-e^{-\beta\omega}}{N\omega}
\frac{
\sum_i g_i e^{-\beta \varepsilon_i}
f_x(\varepsilon_i,\omega,\eta)
}{
\sum_i g_i e^{-\beta \varepsilon_i}
}.
\label{eq:sigma_from_fx_avg_main}
\end{align}
Since $(1-e^{-\beta\omega})/\omega\to\beta$ as $\omega\to0$, defining $f_{x}(\varepsilon_i, 0, \eta)\equiv\lim_{\omega\to0}f_x(\varepsilon_i,\omega,\eta)$, we obtain the dc resistivity
\begin{align}
\rho_x(T)
&=
NT
\frac{\sum_i g_i e^{-\beta \varepsilon_i}}
{\sum_i g_i e^{-\beta \varepsilon_i}f_{x}(\varepsilon_i, 0, \eta)}.
\label{eq:rho_level_resolved_main}
\end{align}
We note that we take the $\omega\to0$ limit directly, neglecting the possible spurious finite-size Drude weight.
As proved in Sec.~SI of SM~\cite{SupplementalMaterials}, for a temperature-independent broadening $\eta$, the dc resistivity [Eq.~\eqref{eq:rho_level_resolved_main}] is strictly linear
in $T$, i.e., $\rho_x(T)=\alpha T$ for all $T>0$ with a
temperature-independent coefficient $\alpha$ and hence zero
residual resistivity, if and only if
$f_{x}(\varepsilon_i, 0, \eta)=f_0$ for every level $\varepsilon_i$.
The two thermal sums in
Eq.~\eqref{eq:rho_level_resolved_main} then cancel,
yielding $\alpha=N/f_0$.

\textit{Numerically extrapolated dc $f_x$ function.---}For the infinite-$U$ Hubbard model [Eq.~\eqref{eq:main_H}] under TBCs, we first evaluate Eq.~\eqref{eq:f_function} at fixed twist $\boldsymbol{\theta}$, replacing the average over degenerate eigenstates with an average over eigenstates within energy bins (Sec.~SIII of SM~\cite{SupplementalMaterials}). This gives $f_{x,\boldsymbol{\theta}}(\varepsilon, \omega, \eta)$, where $\varepsilon$ is the bin center. We note that $\varepsilon$ is the dimensionless energy $(E-E_{\mathrm{GS}})/t$. We then average over twists,
$f_{x}(\varepsilon,\omega,\eta)
=N_\theta^{-1}\sum_{\boldsymbol{\theta}}
f_{x,\boldsymbol{\theta}}(\varepsilon,\omega,\eta)$,
and extrapolate to zero frequency using the same finite-frequency window as
for $\sigma_{\mathrm{reg}}^{xx}$ (Sec.~SIII of SM~\cite{SupplementalMaterials}).
For brevity, $f_x(\varepsilon)$ denotes the intercept of this
finite-frequency quadratic fit. Figure~\ref{fig:main_fx} shows
$f_{x}(\varepsilon)$ for $N=16$ and $18$. It is approximately constant over the central region
$\varepsilon_{f_x}\leq\varepsilon\leq W-\varepsilon_{f_x}$, with $W=8$,
and varies more strongly toward the edges of the spectrum.
The edge scales are
$\varepsilon_{f_x}\approx0.99$ for $N=16$ and $1.11$ for $N=18$
(Sec.~SIII of SM~\cite{SupplementalMaterials}).

The twist-averaged density of states (DOS) $D(\varepsilon)$ for $N=16,18$ [Figs.~\ref{fig:main_fx}(c) and \ref{fig:main_fx}(d)] also has an approximately flat region that extends closer to the spectral edges than that of $f_{x}(\varepsilon)$ (Sec.~SIII of SM~\cite{SupplementalMaterials}). 
Motivated by Eq.~\eqref{eq:rho_level_resolved_main}, we consider 
\begin{align}
\rho_x(T)&=\frac{NT}{\overline{f}_x(T)},
\label{eq:main_model_dcrho}\\
\overline{f}_x(T)&=
\frac{\displaystyle\int_0^W d\varepsilon\,
D(\varepsilon)f_x(\varepsilon)e^{-\varepsilon/\widetilde{T}}}
{\displaystyle\int_0^W d\varepsilon\,
D(\varepsilon)e^{-\varepsilon/\widetilde{T}}},
\label{eq:main_thermal_average_fx}
\end{align}
where $\overline{f}_x(T)$ is the thermal average of $f_x(\varepsilon)$ and
$\widetilde{T}\equiv T/t$. We note that Eq.~\eqref{eq:main_model_dcrho} does
not exactly reproduce the dc resistivity obtained from the twist-averaged dc
conductivity, but for the infinite-$U$ Hubbard model with $N=16$ and $18$ the
two are close for $T/t\gtrsim0.1$ (Sec.~SIII of SM~\cite{SupplementalMaterials}).
To make the connection between $\varepsilon_{f_x}$ and the deviation from strictly linear-in-$T$ resistivity explicit, we further approximate 
$f_x(\varepsilon)$ by a symmetric ramp--plateau--ramp profile (see Fig.~\ref{fig:main_fx}):
\begin{equation}
f_x(\varepsilon)=
\begin{cases}
f_0\,\varepsilon/\varepsilon_{f_x},
& 0\leq\varepsilon<\varepsilon_{f_x},\\
f_0,
& \varepsilon_{f_x}\leq\varepsilon
\leq W-\varepsilon_{f_x},\\
f_0\,(W-\varepsilon)/\varepsilon_{f_x},
& W-\varepsilon_{f_x}<\varepsilon\leq W,
\end{cases}
\label{eq:main_toy}
\end{equation}
where $f_0$ is the plateau value. We further approximate the DOS by a constant, $D(\varepsilon)=D_0$, which then cancels between the numerator and denominator of Eq.~\eqref{eq:main_thermal_average_fx}. The effects of these approximations are discussed in Sec.~SIII of SM~\cite{SupplementalMaterials}.

With these approximations, $\rho_x(T)$ has a closed form (Sec.~SIII of SM~\cite{SupplementalMaterials}). A high-temperature expansion gives $\rho_x(T)$ and its local slope $a(T) = d \rho_x / d(T/t)$:
\begin{align}
\rho_x(T)
&=
a_\infty \frac{T}{t}
\left[
1+\left(\frac{T_{\mathrm{cr}}}{T}\right)^2
+\mathcal{O}\!\left((t/T)^4\right)
\right],
\nonumber\\
a(T)
&=
a_\infty
\left[
1-\left(\frac{T_{\mathrm{cr}}}{T}\right)^2
+\mathcal{O}\!\left((t/T)^4\right)
\right],
\label{eq:main_highT_rhox_a}
\end{align}
where
\begin{align}
a_\infty=
\frac{NWt}{f_0(W-\varepsilon_{f_x})}, \quad
T_{\mathrm{cr}}/t=
\sqrt{
\frac{\varepsilon_{f_x}
(W-\varepsilon_{f_x})}{12}
}.
\label{eq:main_Tcr}
\end{align}
Here $a_\infty=a(T\to\infty)$ is the
slope in the high-temperature limit. Equation~\eqref{eq:main_highT_rhox_a} shows that $\rho_x(T)$ is linear in $T$ at high temperatures, and that upon cooling, the leading deviations of both $\rho_x(T)$ and $a(T)$ from their high-temperature behavior are controlled by the temperature scale $T_{\mathrm{cr}}$.

\textit{Numerical results for dc resistivity and crossover temperature scale.---} We now turn to the extrapolated dc resistivity $\rho_x(T)$ of the model in Eq.~\eqref{eq:main_H} over $0.05 \leq T/t \leq 16$, obtained from full ED in the single-hole sector with all $S^z$ sectors included (Sec.~SII of SM~\cite{SupplementalMaterials}). Section~SII of SM also resolves the optical resistivity into individual $S^z$ sectors within the single-hole sector for the $4\times4$ cluster under PBCs. Figures~\ref{fig:main_rho}(a) and \ref{fig:main_rho}(b) show $\rho_x(T)$ over $0.05 \leq T/t \leq 1.5$ for $N=16$ and $18$, respectively, and the insets show the full range $0.05 \leq T/t \leq 16$. Figures~\ref{fig:main_rho}(c) and \ref{fig:main_rho}(d) show the corresponding local slope $a(T)$ over $0.05 \leq T/t \leq 8$, in which the blue dashed lines mark the numerically obtained high-temperature slope $a_{\infty}$.

In the high-temperature limit $T\gg t$, all eigenstates carry equal thermal weight and $(1-e^{-\beta\omega})/\omega\to\beta$ in Eq.~\eqref{eq:sigma_regular_with_eta_L_main}, so the resistivity becomes linear in $T$~\cite{PhysRevB.102.115142,PhysRevB.104.165143,PhysRevB.95.041110,PhysRevLett.110.086401,PhysRevB.81.054512,PhysRevB.94.235115}; see the insets of Figs.~\ref{fig:main_rho}(a) and \ref{fig:main_rho}(b). We note that linear-in-$T$ resistivity at high temperatures also arises in conventional metals, from quasiparticle scattering by thermal lattice vibrations in the classical equipartition regime~\cite{BruinScience2013,HartnollMackenzieRMP2022,PhysRevB.108.125145}.
The local slope likewise approaches a constant: at $T/t\approx 7$, $a(T)$ is within approximately $0.4\%$ and $0.5\%$ of its asymptotic value $a_\infty$ (blue dashed lines) for $N=16$ and $18$, respectively. Fitting $a(T)$ to the leading
high-temperature form in Eq.~\eqref{eq:main_highT_rhox_a} over
$3.5\leq T/t\leq8$ (Sec.~SIII of SM~\cite{SupplementalMaterials}), with both $a_\infty$ and
$T_{\mathrm{cr}}$ as free parameters, yields
$T_{\mathrm{cr}}/t\approx0.414$ for $N=16$ and $0.467$ for $N=18$, which are smaller than the values predicted by Eq.~\eqref{eq:main_Tcr} using the ramp--plateau--ramp approximation [Eq.~\eqref{eq:main_toy}] and a constant DOS.

\begin{figure}[t]
\centering
\includegraphics[width=\columnwidth, keepaspectratio]{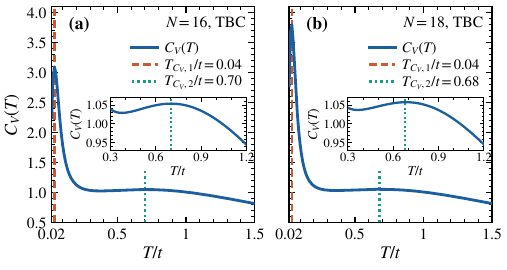}
\caption{(a),(b) Specific heat $C_V(T)$ for $N=16$ and $18$, respectively. Orange dashed lines mark the sharp peak at $T_{C_V,1}/t\approx0.04$, and green dotted lines mark the broad peak at $T_{C_V,2}/t\approx0.70$ and $0.68$, respectively. Insets show a zoom-in over $0.3 \leq T/t \leq 1.2$.}
\label{fig:main_scales}
\end{figure}

Beyond the high-temperature limit, within $T/t \geq 0.05$ we find a gradual crossover between two approximately linear-in-$T$ resistivity regimes, with $T_{\mathrm{cr}}$ [vertical green dotted lines in Figs.~\ref{fig:main_rho}(a) and \ref{fig:main_rho}(b)] serving as the characteristic crossover temperature. In the high-temperature regime, the linearity is only approximate because $a(T)$ still depends on temperature before saturating in the $T/t \to \infty$ limit [Figs.~\ref{fig:main_rho}(c) and \ref{fig:main_rho}(d)]. This departure from high-temperature linearity is related to the edge scale $\varepsilon_{f_x}$ through $T_{\mathrm{cr}}$ [Eqs.~\eqref{eq:main_highT_rhox_a} and~\eqref{eq:main_Tcr}]. Upon further cooling, the resistivity leaves the high-temperature regime [Figs.~\ref{fig:main_rho}(a) and \ref{fig:main_rho}(b)], and a second approximately linear-in-$T$ regime appears at low temperatures. Although $a(T)$ varies more rapidly in this regime than in the high-temperature regime, a linear fit over $0.05 \leq T/t \leq T_{\mathrm{cr}}/t - 0.15$ describes the data well, with $R^{2} \approx 0.998$ for both $N=16$ and $18$. Because the crossover is gradual, we set the upper bound of the fitting window to $T_{\mathrm{cr}}/t-0.15$ rather than $T_{\mathrm{cr}}/t$. The insets of Figs.~\ref{fig:main_rho}(a) and \ref{fig:main_rho}(b) compare this linear fit with the high-temperature linear behavior $a_\infty T/t$.

In addition to the dc resistivity, we calculate the specific heat
$C_V(T)=\beta^2(\langle E^2\rangle-\langle E\rangle^2)$, evaluated at each
twist and then averaged over twists (Sec.~SIII of SM~\cite{SupplementalMaterials}). Figures~\ref{fig:main_scales}(a)
and~\ref{fig:main_scales}(b) show $C_V(T)$ for $N=16$ and $18$, respectively.
For both sizes, $C_V(T)$ exhibits two peaks.
The sharp peak at
$T_{C_V,1}/t\approx0.04$ lies below the temperature range of our
transport calculations. This peak is consistent with the crossover from a paramagnet to a regime with finite ferromagnetic correlations reported in Ref.~\cite{fratini2026strangemetaltransportcoupling}. A second, broader peak appears at
$T_{C_V,2}/t\approx0.70$ for $N=16$ and $0.68$ for $N=18$ [insets of Figs.~\ref{fig:main_scales}(a) and \ref{fig:main_scales}(b)]. Its position is comparable to, but does not coincide with, the fitted
$T_{\mathrm{cr}}$, which is consistent with a gradual crossover. Observable-dependent crossover scales, rather
than a single sharp transition temperature, have also been reported in
Ref.~\cite{relatedwork}.

In the thermodynamic limit, the model [Eq.~\eqref{eq:main_H}] with a single doped hole is paramagnetic for all $T>0$~\cite{PhysRevB.43.8779}, so that $T_{C_V,1}/t \to 0$. It remains open whether, in this limit, the approximately linear-in-$T$ low-temperature regime extends to zero temperature with finite $T_{\mathrm{cr}}/t$, or vanishes with $T_{\mathrm{cr}}/t \to 0$. This question cannot be settled with the system sizes currently accessible to ED.

\textit{Discussion and outlook.---}We have proved that a constant energy-resolved current-response function $f_x$ is necessary and sufficient for
$\rho_x(T)=\alpha T$ at all $T>0$ with a finite, temperature-independent $\alpha$. For the infinite-$U$ Hubbard model [Eq.~\eqref{eq:main_H}] with a single doped hole on $N=16$ and $18$ square clusters, our ED results show that $f_x(\varepsilon)$ has a broad, approximately flat central
region in the many-body spectrum and pronounced energy dependence within $\varepsilon_{f_x}$ of the spectral edges. We have further shown that this energy scale is related to $T_{\mathrm{cr}}$, which characterizes the deviation of the resistivity from its high-temperature behavior. Our ED results also identify two approximately linear-in-$T$
resistivity regimes with a gradual crossover between them (Fig.~\ref{fig:main_rho}); the crossover temperatures are $T_{\mathrm{cr}}/t\approx0.414$ and $0.467$ for $N=16$ and $18$, respectively.

Several questions remain open. We have calculated the energy-resolved level statistics of Eq.~\eqref{eq:main_H}, which follow the Gaussian orthogonal ensemble (GOE) prediction away from the spectral edges (Sec.~SI of SM~\cite{SupplementalMaterials}). How this connects to the central plateau in $f_x(\varepsilon)$ within the eigenstate thermalization
hypothesis~\cite{PhysRevA.43.2046, PhysRevE.50.888,
rigolThermalizationItsMechanism2008c, D'Alessio03052016,
PhysRevB.73.035113} remains to be clarified. It would also be interesting to identify which properties of a Hamiltonian, such as the infinite-$U$ constraint or the single-hole sector, reduce $\varepsilon_{f_x}$ and thereby widen the plateau of $f_x(\varepsilon)$. Another question, raised above, is whether the low-temperature linear-in-$T$ resistivity regime survives in the thermodynamic limit. Finally, our analysis of $f_x(\varepsilon)$ can be generalized to other models, including the
recently developed momentum-mixing variant of the Hatsugai--Kohmoto
model~\cite{phillipsExactTheorySuperconductivity2020, maiTwistingHubbardModel2026, ry4h-kkvs, PhysRevB.111.045126, ma2025charge}, and may help identify models that exhibit linear-in-$T$ resistivity down to lower temperatures.

\textit{Note added.---}While completing this manuscript, we became aware of
Ref.~\cite{relatedwork}, which studies the same model and
also identifies two regimes of approximately
linear-in-$T$ resistivity with different slopes. The authors find
that the charge compressibility scales as $\chi_c\sim1/T$
at high temperatures and approaches a nearly constant value
at low temperatures. Using, among other diagnostics, the departure
of $1/\chi_c$ from its high-temperature form, they identify an
observable-dependent crossover scale
$T_{\mathrm{quant}}/t\approx0.5$--$0.8$. Our crossover temperatures,
$T_{\mathrm{cr}}/t\approx0.414$ ($N=16$) and $0.467$ ($N=18$), are
comparable in magnitude to $T_{\mathrm{quant}}$.

\textit{Acknowledgments.---}We thank David A. Huse and A. Patel for insightful discussions on the necessary and sufficient condition for strictly linear-in-$T$ resistivity. H.J.C. also thanks S. Sachdev and V. Oganesyan for discussions on the subject of $T$-linear resistivity, and M. Stoudenmire and Flatiron Institute for their hospitality. S.L. and B.A.B. were supported by the Gordon and Betty Moore Foundation (grant GBMF8685) towards the Princeton theory program, the Gordon and Betty Moore Foundation's EPiQS Initiative (grant GBMF11070), the Global Collaborative Network Grant at Princeton University, the Simons Investigator Grant No. 404513, the NSF MRSEC (grant DMR-2011750), the Simons Collaboration on New Frontiers in Superconductivity (grant SFI-MPS-NFS-00006741-01), the Princeton Catalysis Initiative (PCI), the Schmidt Transformative Technology Fund at Princeton University and the NSF through the AI Research Institutes (program award DMR-2433348). Y. M. and P. W. P. acknowledge funding
from UIUC Research Board, Award No. RB26125. H.J.C. was supported by NSF grant DMR-2046570 and DOE grant DE-SC0026488 (Subaward: SUB00001040). A.K. was supported through a Dirac postdoctoral fellowship at the National High Magnetic Field Laboratory (NHMFL). The NHMFL is supported by the National Science Foundation through NSF/DMR-2128556 and the State of Florida. We thank the Research Computing Center at Florida State University for computational resources.
We also acknowledge the use of Bridges-2 at the Pittsburgh Supercomputing Center (PSC) through allocation PHY260268: ``Unconventional Metallicity and Non-Fermi-Liquid Behavior in Clean Strongly Correlated Models'' from the Advanced Cyberinfrastructure Coordination Ecosystem: Services and Support (ACCESS) program, which is supported by U.S. National Science Foundation grants $\#$2138259, $\#$2138286, $\#$2138307, $\#$2137603, and $\#$2138296. The authors used OpenAI Codex (GPT-6 Astra) and Claude Code (Opus 5) for manuscript revision.

\bibliographystyle{apsreve}
\let\oldaddcontentsline\addcontentsline
\renewcommand{\addcontentsline}[3]{}
\bibliography{ref}
\let\addcontentsline\oldaddcontentsline

\clearpage
\onecolumngrid

\begin{center}
\textbf{\large Supplemental Material for ``Energy-resolved current response and the crossover between two approximately linear-in-$T$ resistivity regimes in the infinite-$U$ Hubbard model''}
\end{center}

\renewcommand{\thesection}{S\Roman{section}}
\renewcommand{\theHsection}{S\arabic{section}}
\renewcommand{\theequation}{\thesection.\arabic{equation}}
\renewcommand{\theHequation}{S\arabic{section}.\arabic{equation}}
\renewcommand{\thefigure}{S\arabic{figure}}
\renewcommand{\theHfigure}{S\arabic{figure}}
\renewcommand{\thetable}{S\arabic{table}}
\renewcommand{\theHtable}{S\arabic{table}}
\renewcommand{\thepage}{S\arabic{page}}
\setcounter{section}{0}
\setcounter{equation}{0}
\setcounter{figure}{0}
\setcounter{table}{0}
\setcounter{page}{1}
\setcounter{secnumdepth}{4}
\newtheorem{theorem}{Theorem}

\makeatletter
\def\tocmax@section{2.5em}
\def\tocmax@subsection{2.5em}
\def\tocmax@subsubsection{2.5em}
\makeatother

\tableofcontents

\section{Model, spectral statistics, and transport formalism}
\label{sec:model_and_transport_formalism}

This section collects the model definitions and the transport formalism used
throughout this work. In
Sec.~\ref{sub:hamiltonan_of_infinite_u_hubbard_model}, we introduce the
Hamiltonian of the infinite-$U$ Hubbard model on the conventional $L\times L$
and the rotated $18$-site square clusters. We also specify the periodic and
twisted boundary conditions, the twist grid used in the calculations, and the symmetries of the model. In
Sec.~\ref{sub:level_statistics}, we present the energy-resolved level
statistics of the infinite-$U$ Hubbard model. These level statistics indicate that the infinite-$U$ Hubbard model is nonintegrable. In
Sec.~\ref{sec:Kubo_formula_and_optical_conductivity}, we define the
$x$-direction current operator, write down the Kubo formula for the optical
conductivity, and describe the procedure used to extract the dc conductivity
at finite size. In Sec.~\ref{sub:analytical_optical_conductivity_with_single_hole_in_fully_polarized_background}, we present analytical results for the optical conductivity of the infinite-$U$ Hubbard model with a single doped hole in a fully polarized background. This exactly solvable case also illustrates why the order of limits in Eq.~\eqref{eq:order_of_limits} matters. In Sec.~\ref{sec:definition_of_f_function}, we define, following Ref.~\cite{PhysRevB.105.L201108}, the energy-resolved current-response function $f_x$ and prove
that, for an arbitrary Hamiltonian, the resistivity is strictly linear in $T$ if
and only if the dc $f_x$ is independent of energy.

\subsection{\texorpdfstring{Infinite-$U$ Hubbard model: clusters, boundary conditions, and symmetries}{Infinite-U Hubbard model: clusters, boundary conditions, and symmetries}}
\label{sub:hamiltonan_of_infinite_u_hubbard_model}

We consider the Hubbard model in the infinite-$U$ limit on a square
lattice with $N$ sites. The infinite on-site repulsion excludes double
occupancy, which we implement through the projected fermion operators
\begin{align}
\tilde{c}^{\dagger}_{\mathbf{r},\sigma}
&=\left(1-n_{\mathbf{r},\bar{\sigma}}\right)c^{\dagger}_{\mathbf{r},\sigma},
\notag\\
\tilde{c}_{\mathbf{r},\sigma}
&=\left(1-n_{\mathbf{r},\bar{\sigma}}\right)c_{\mathbf{r},\sigma},
\label{eq:projected_operators}
\end{align}
where $c^{\dagger}_{\mathbf{r},\sigma}$ ($c_{\mathbf{r},\sigma}$) creates
(annihilates) a fermion of spin $\sigma=\uparrow,\downarrow$ at site
$\mathbf{r}$,
$n_{\mathbf{r},\sigma}=c^{\dagger}_{\mathbf{r},\sigma}c_{\mathbf{r},\sigma}$,
and $\bar{\sigma}$ denotes the spin opposite to $\sigma$. The Hamiltonian of
the infinite-$U$ Hubbard model is given by
\begin{align}
\hat{H} =
-t \sum_{\sigma=\uparrow,\downarrow}
\sum_{\langle \mathbf{r},\mathbf{r}' \rangle}
\left(
\tilde{c}^{\dagger}_{\mathbf{r},\sigma}\tilde{c}_{\mathbf{r}',\sigma}
+\tilde{c}^{\dagger}_{\mathbf{r}',\sigma}\tilde{c}_{\mathbf{r},\sigma}
\right),
\label{eq:H2DNN}
\end{align}
where $\langle \mathbf{r},\mathbf{r}' \rangle$ runs over nearest-neighbor
bonds, each counted once, and $t$ is the hopping amplitude.

\begin{figure}[t]
\centering
\includegraphics[width=3.375in]{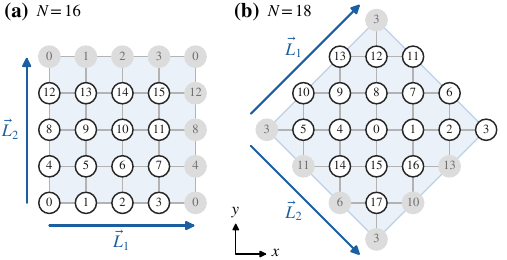}
\caption{Schematics of the two clusters used in this work. (a) The $4\times4$
square cluster ($N=16$) with $(\vec{L}_{1}, \vec{L}_{2}) = (4\hat{x}, 4\hat{y})$.
(b) The rotated $18$-site square cluster with
$(\vec{L}_{1}, \vec{L}_{2}) = (3 \hat{x} + 3 \hat{y}, 3 \hat{x} - 3 \hat{y})$.
Open circles denote physical sites, labeled by the site indices used in the ED
calculations; gray circles denote periodic images introduced by the boundary
conditions, labeled by the corresponding physical sites. Gray lines denote
nearest-neighbor bonds, and the shaded regions are the cells spanned by
$\vec{L}_{1}$ and $\vec{L}_{2}$.}
\label{fig:Rotated_Lattice_N18}
\end{figure}

We consider two square clusters. The first is the conventional $L \times L$ cluster with $L=4$, i.e., $N=L^2=16$ sites (see Fig.~\ref{fig:Rotated_Lattice_N18}(a)), on which Eq.~\eqref{eq:H2DNN} with periodic
boundary conditions (PBCs) reads
\begin{align}
\hat{H} &=
-t \sum_{\sigma=\uparrow,\downarrow}\sum_{x,y=0}^{L-1}
\left(
\tilde{c}^{\dagger}_{x+1,y,\sigma}\tilde{c}_{x,y,\sigma}
+\tilde{c}^{\dagger}_{x,y,\sigma}\tilde{c}_{x+1,y,\sigma}
\right)
\notag\\
&\quad
-t \sum_{\sigma=\uparrow,\downarrow}\sum_{x,y=0}^{L-1}
\left(
\tilde{c}^{\dagger}_{x,y+1,\sigma}\tilde{c}_{x,y,\sigma}
+\tilde{c}^{\dagger}_{x,y,\sigma}\tilde{c}_{x,y+1,\sigma}
\right),
\label{eq:H2D}
\end{align}
where $\mathbf{r}=(x,y)$ and all site indices are understood modulo $L$, i.e.,
$\tilde{c}^{(\dagger)}_{x+L,y,\sigma}
=\tilde{c}^{(\dagger)}_{x,y+L,\sigma}
=\tilde{c}^{(\dagger)}_{x,y,\sigma}$. The second is the rotated $18$-site square cluster shown in Fig.~\ref{fig:Rotated_Lattice_N18}(b).

At half filling, corresponding to one fermion per site, the infinite-$U$ Hubbard model is a Mott insulator: no hopping is allowed and all
spin configurations are degenerate. With a single doped hole and PBCs, the ground state
of this model is known exactly: it is the Nagaoka ferromagnetic
state~\cite{PhysRev.147.392}, with ground-state energy $E_{\mathrm{GS}}=-4t$ (see also Sec.~\ref{sub:analytical_optical_conductivity_with_single_hole_in_fully_polarized_background}).
Throughout this work, we consider a single doped hole, so that the hole density
is $1/N$. Because of the limited system sizes accessible to ED, we do not consider a fixed finite hole density. It therefore remains unclear whether a finite hole density leads to charge-transport behavior different from that of the single-hole case, as is the case in the one-dimensional infinite-$U$ Hubbard model~\cite{liu2026transporttemperature1exact}. For both PBCs and the twisted boundary conditions (TBCs) introduced below [Eq.~\eqref{eq:twist_bc}], we take the PBC ground-state energy $E_{\mathrm{GS}}=-4t$ as a common reference for all twists and define the dimensionless energy $\varepsilon = (E-E_{\mathrm{GS}})/t$.

In the numerical calculations of charge transport, we employ TBCs to reduce finite-size effects. Let $\vec{L}_1$ and
$\vec{L}_2$ denote the two periodicity vectors of the cluster, and let $\hat{x}$ and $\hat{y}$ be the nearest-neighbor unit vectors. For a given twist
$\boldsymbol{\theta}=(\theta_1,\theta_2)$, with each boundary phase defined modulo $2\pi$, the TBCs are
imposed through
\begin{equation}
\tilde{c}_{\mathbf{r}+\vec{L}_\mu,\sigma}
=e^{i\theta_\mu}\,\tilde{c}_{\mathbf{r},\sigma},
\qquad \mu=1,2 ,
\label{eq:twist_bc}
\end{equation}
with $\boldsymbol{\theta}=\mathbf{0}$ recovering PBCs. It is convenient to
distribute the twist uniformly over all bonds, which amounts to introducing a
uniform vector potential
$\mathbf{A}_{\boldsymbol{\theta}}=A_x\hat{x}+A_y\hat{y}$ determined by
\begin{equation}
\mathbf{A}_{\boldsymbol{\theta}}\cdot\vec{L}_\mu=\theta_\mu,
\qquad \mu=1,2 .
\label{eq:A_definition}
\end{equation}
For the $L\times L$ cluster with $(\vec{L}_{1}, \vec{L}_{2}) = (L \hat{x}, L \hat{y})$, Eq.~\eqref{eq:A_definition} gives
\begin{equation}
A_x=\frac{\theta_1}{L},
\qquad
A_y=\frac{\theta_2}{L},
\label{eq:square_AxAy}
\end{equation}
while for the $18$-site cluster [Fig.~\ref{fig:Rotated_Lattice_N18}(b)] with $(\vec{L}_{1}, \vec{L}_{2}) = (3 \hat{x} + 3 \hat{y}, 3 \hat{x} - 3 \hat{y})$, it yields
\begin{align}
A_x=\frac{\theta_1+\theta_2}{6},
\qquad
A_y=\frac{\theta_1-\theta_2}{6}.
\label{eq:18square_AxAy}
\end{align}
With this convention [Eqs.~\eqref{eq:square_AxAy} and~\eqref{eq:18square_AxAy}], the Hamiltonian of the infinite-$U$ Hubbard model under TBCs takes the same form for both clusters,
\begin{align}
\hat{H}(\boldsymbol{\theta})
&= -t \sum_{\sigma=\uparrow,\downarrow}\sum_{\mathbf{r}}
\left[
e^{iA_x}\tilde{c}^{\dagger}_{\mathbf{r},\sigma}
\tilde{c}_{\mathbf{r}+\hat{x},\sigma}
+ e^{-iA_x}\tilde{c}^{\dagger}_{\mathbf{r}+\hat{x},\sigma}
\tilde{c}_{\mathbf{r},\sigma}
\right]
\notag\\
&\quad
-t \sum_{\sigma=\uparrow,\downarrow}\sum_{\mathbf{r}}
\left[
e^{iA_y}\tilde{c}^{\dagger}_{\mathbf{r},\sigma}
\tilde{c}_{\mathbf{r}+\hat{y},\sigma}
+ e^{-iA_y}\tilde{c}^{\dagger}_{\mathbf{r}+\hat{y},\sigma}
\tilde{c}_{\mathbf{r},\sigma}
\right].
\label{eq:H2D_TBCs}
\end{align}
Since $\mathbf{A}_{\boldsymbol{\theta}}$ is uniform, no flux threads any
plaquette, and the Hamiltonian remains translation invariant for every
$\boldsymbol{\theta}$.

Specifically, we use $N_\theta=121$ equally weighted twists, parametrized by
\begin{equation}
\phi_x=\frac{2\pi i_x}{11},\qquad
\phi_y=\frac{2\pi i_y}{11},\qquad i_x,i_y=0,1,\ldots,10.
\label{eq:actual_twist_grid}
\end{equation}
For the $N=16$ and $18$ clusters, we set
\begin{equation}
(\theta_1,\theta_2)=
\begin{cases}
(\phi_x,\phi_y), & N=16,\\[4pt]
\left(\dfrac{\phi_x+\phi_y}{2},\dfrac{\phi_x-\phi_y}{2}\right) \pmod{2\pi}, & N=18,
\end{cases}
\label{eq:actual_rotated_boundary_twists}
\end{equation}
so that Eqs.~\eqref{eq:square_AxAy} and~\eqref{eq:18square_AxAy} give
\begin{equation}
(A_x,A_y)=
\begin{cases}
(\phi_x/4,\phi_y/4),&N=16,\\
(\phi_x/6,\phi_y/6),&N=18.
\end{cases}
\label{eq:actual_twist_bond_phases}
\end{equation}
Exchanging $x$ and $y$ maps
$(i_x,i_y)$ to $(i_y,i_x)$, so this twist grid is invariant under $x\leftrightarrow y$.

For both boundary conditions and both cluster geometries ($N=16,18$), the infinite-$U$ Hubbard model conserves the
total charge and the total spin, i.e., it has a $U(1)\times SU(2)$ symmetry,
and it is invariant under lattice translations. In the transport ED calculations in Sec.~\ref{sec:numerical_results}, we
resolve the $U(1)\times U(1)$ subgroup generated by the particle number and by
$S^z$, together with translations. For the level statistics in
Sec.~\ref{sub:level_statistics}, we further resolve the total spin and the
remaining spatial symmetries at each twist.

To exploit translation symmetry in the ED calculation, we list the
allowed momenta on the rotated $18$-site cluster (see Fig.~\ref{fig:Rotated_Lattice_N18}(b)). They satisfy
$e^{i\vec{k}\cdot\vec{L}_1}=e^{i\vec{k}\cdot\vec{L}_2}=1$, where $(\vec{L}_{1}, \vec{L}_{2}) = (3 \hat{x} + 3 \hat{y}, 3 \hat{x} - 3 \hat{y})$, i.e.,
$3(k_x+k_y)=2\pi m$ and $3(k_x-k_y)=2\pi n$ with $m,n\in\mathbb{Z}$, or
equivalently
\begin{equation}
k_x=\frac{(m+n)\pi}{3}\equiv\frac{m_x\pi}{3},
\qquad
k_y=\frac{(m-n)\pi}{3}\equiv\frac{m_y\pi}{3}.
\label{eq:allowed_momenta_rotated}
\end{equation}
Because $m_x+m_y=2m$ is even, the allowed pairs in the first Brillouin zone are
those with $m_x,m_y\in\{0,1,\ldots,5\}$ and $m_x+m_y$ even,
\begin{align}
(m_x,m_y)=&(0,0),(0,2),(0,4),(1,1),(1,3),(1,5),
\notag\\
&(2,0),(2,2),(2,4),(3,1),(3,3),(3,5),
\notag\\
&(4,0),(4,2),(4,4),(5,1),(5,3),(5,5),
\label{eq:momentum_list_rotated}
\end{align}
which are $18$ momentum pairs, consistent with the $18$ sites of the cluster.

\subsection{Energy-resolved level statistics}
\label{sub:level_statistics}

Before turning to charge transport, we use the energy-resolved level statistics to assess whether the infinite-$U$ Hubbard model [Eq.~\eqref{eq:H2D_TBCs}] is nonintegrable. For the $N=16$- and $18$-site clusters, we compute these statistics at all $121$ twists specified by
Eqs.~\eqref{eq:actual_twist_grid} and~\eqref{eq:actual_rotated_boundary_twists}. To probe the intrinsic level statistics, all symmetries must be resolved and the statistics computed within each symmetry sector. We resolve particle-number conservation, the $SU(2)$ spin symmetry, and translations along the $x$ and $y$ directions, and label the resulting blocks by $(N_h=1,S,S^z,k_x,k_y)$, where $N_h$ is the number of holes. We note that, at some twists, these blocks can be further reduced by additional spatial symmetries, which we also take into account. For example, for $N=18$ with PBCs, the sectors with $(k_x,k_y)=(0,\pm2\pi/3)$ are further split by the reflection $(x,y)\mapsto(-x,y)$.

Let $b$ label such a fully reduced block at twist
$\boldsymbol{\theta}$, with dimension $D_{b}$ and eigenvalues
$E_{1,b}(\boldsymbol{\theta})\leq\cdots\leq E_{D_b,b}(\boldsymbol{\theta})$.
We use the level-spacing ratio~\cite{PhysRevB.75.155111,PhysRevLett.110.084101,PhysRevX.12.011006}
\begin{equation}
\delta_{n,b}^{\boldsymbol{\theta}}
=E_{n+1,b}(\boldsymbol{\theta})-E_{n,b}(\boldsymbol{\theta}),\qquad
r_{n,b}^{\boldsymbol{\theta}}
=\frac{\min(\delta_{n-1,b}^{\boldsymbol{\theta}},\delta_{n,b}^{\boldsymbol{\theta}})}
{\max(\delta_{n-1,b}^{\boldsymbol{\theta}},\delta_{n,b}^{\boldsymbol{\theta}})},
\quad n=2,\ldots,D_b-1.
\label{eq:level_spacing_ratio}
\end{equation}
In terms of the dimensionless energy $\varepsilon = (E+4t)/t$ defined in
Sec.~\ref{sub:hamiltonan_of_infinite_u_hubbard_model}, we choose the energy bins
such that the band center $E=0$ ($\varepsilon=4$) lies at the center of
the bin $[3.9,4.1)$. Extending this grid in both
directions and truncating it at $\varepsilon=0$ and $8$ gives $39$ interior
bins of width $0.2$, from $[0.1,0.3)$ to $[7.7,7.9)$ with centers
$0.2,0.4,\ldots,7.8$, and two edge bins, $[0,0.1)$ and $[7.9,8]$, of width
$0.1$ with centers $0.05$ and $7.95$.

For each bin $B_j$, block $b$, and twist $\boldsymbol{\theta}$, we evaluate $r_{n,b}^{\boldsymbol{\theta}}$ [Eq.~\eqref{eq:level_spacing_ratio}] only when the energies of the three levels $n-1$, $n$, and $n+1$ all lie in $B_j$. Let $\mathcal I_{\boldsymbol{\theta},b,j}$ denote the set of such indices $n$ of $r_{n,b}^{\boldsymbol{\theta}}$, and let
$m^{\boldsymbol{\theta}}_{b,j}=|\mathcal I_{\boldsymbol{\theta},b,j}|$ be the number of ratios obtained. The average ratio in $B_j$ at fixed twist $\boldsymbol{\theta}$ is
\begin{equation}
\overline r^{\boldsymbol{\theta}}_{j}
=\frac{\displaystyle\sum_b\sum_{n\in\mathcal I_{\boldsymbol{\theta},b,j}}
r_{n,b}^{\boldsymbol{\theta}}}
{m^{\boldsymbol{\theta}}_{j}},\qquad
m^{\boldsymbol{\theta}}_{j}=\sum_b m^{\boldsymbol{\theta}}_{b,j},
\label{eq:level_ratio_within_twist}
\end{equation}
for $m^{\boldsymbol{\theta}}_{j}>0$, and its equal-weight average over twists is
\begin{equation}
\overline r_j=\frac{1}{K_j}
\sum_{\boldsymbol{\theta}\in\mathcal T_j}
\overline r^{\boldsymbol{\theta}}_{j},
\label{eq:level_ratio_twist_average}
\end{equation}
where
$\mathcal T_j=\{\boldsymbol{\theta}:m^{\boldsymbol{\theta}}_{j}>0\}$ and
$K_j=|\mathcal T_j|$.

\begin{figure}[htbp]
\centering
\includegraphics[width=5.375in]{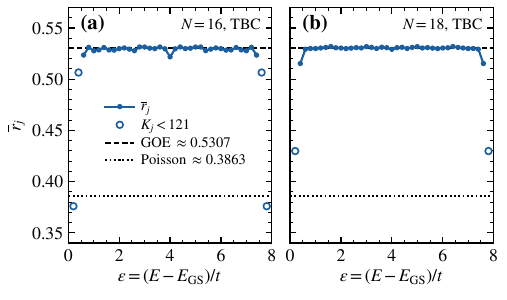}
\caption{Energy-resolved level statistics $\overline r_j$
[Eq.~\eqref{eq:level_ratio_twist_average}] for (a) $N=16$ and (b) $N=18$. Filled and open points have $K_j=121$ and $K_{j}<121$, respectively. Black dashed and dotted lines mark the GOE and Poisson values, $\langle r\rangle_{\mathrm{GOE}}\simeq0.5307$ and $\langle r\rangle_{\mathrm{P}}=2\ln2-1\simeq0.3863$, respectively.}
\label{fig:level_statistics_twist_average}
\end{figure}

Figure~\ref{fig:level_statistics_twist_average} shows $\overline r_j$ [Eq.~\eqref{eq:level_ratio_twist_average}] for the $N=16$ and $18$ clusters. 
For bin centers $0.8\leq\varepsilon\leq7.2$, $\overline r_j$ is close to the Gaussian orthogonal ensemble (GOE) value $\langle r\rangle_{\mathrm{GOE}}\simeq0.5307$~\cite{PhysRevLett.110.084101}.
Near the spectral edges, $K_{j}<121$ and $\overline r_j$ falls below the GOE value. These energy-resolved level statistics therefore indicate that the infinite-$U$ Hubbard model is nonintegrable.

\subsection{Current operator, Kubo formula, and extraction of the dc conductivity}
\label{sec:Kubo_formula_and_optical_conductivity}

Having specified the model (Sec.~\ref{sub:hamiltonan_of_infinite_u_hubbard_model}) and its level statistics (Sec.~\ref{sub:level_statistics}), we now introduce the
current operator and the Kubo formula for the optical conductivity of finite
clusters, and describe how the dc limit is extracted.

The $x$-direction current operator follows from the vector potential introduced
in Eq.~\eqref{eq:A_definition},
\begin{align}
\hat{J}^{x}(\boldsymbol{\theta})
=-\frac{\partial \hat{H}(\boldsymbol{\theta})}{\partial A_x}
= it \sum_{\sigma=\uparrow,\downarrow}\sum_{\mathbf{r}}
\left[
e^{iA_x}\tilde{c}^{\dagger}_{\mathbf{r},\sigma}
\tilde{c}_{\mathbf{r}+\hat{x},\sigma}
- e^{-iA_x}\tilde{c}^{\dagger}_{\mathbf{r}+\hat{x},\sigma}
\tilde{c}_{\mathbf{r},\sigma}
\right],
\label{eq:Jx_TBCs}
\end{align}
which is equivalent to the expression obtained from the lattice continuity
equation. At $\boldsymbol{\theta}=\mathbf{0}$, it reduces to the PBC
current operator
\begin{align}
\hat{J}^{x}
= it \sum_{\sigma=\uparrow,\downarrow}\sum_{\mathbf{r}}
\left(
\tilde{c}^{\dagger}_{\mathbf{r},\sigma}\tilde{c}_{\mathbf{r}+\hat{x},\sigma}
-\tilde{c}^{\dagger}_{\mathbf{r}+\hat{x},\sigma}\tilde{c}_{\mathbf{r},\sigma}
\right).
\label{eq:2DxCurrent}
\end{align}
For the twist grid of Eqs.~\eqref{eq:actual_twist_grid}--\eqref{eq:actual_twist_bond_phases}, the twist-averaged charge transport is equivalent along the $x$ and $y$ directions, so we focus on transport along $x$ throughout. We also note that the Nagaoka ferromagnetic ground state~\cite{PhysRev.147.392} is an eigenstate of the current operator in Eq.~\eqref{eq:2DxCurrent} with
eigenvalue zero (see also Sec.~\ref{sub:analytical_optical_conductivity_with_single_hole_in_fully_polarized_background}).

At each twist $\boldsymbol{\theta}$, we denote the eigenstates of
$\hat{H}(\boldsymbol{\theta})$ [Eq.~\eqref{eq:H2D_TBCs}] by $\ket{n(\boldsymbol{\theta})}$, their energies by $E_n(\boldsymbol{\theta})$, and the current matrix elements by
$J^x_{mn}(\boldsymbol{\theta})=\bra{m(\boldsymbol{\theta})}\hat{J}^x(\boldsymbol{\theta})\ket{n(\boldsymbol{\theta})}$.
The regular part of the longitudinal optical conductivity then reads
\begin{equation}
\sigma^{xx}_{\mathrm{reg},\boldsymbol{\theta}}(\omega,T)
=\pi \frac{1-e^{-\beta\omega}}{NZ(\boldsymbol{\theta})\omega}
\sum_{m,n} e^{-\beta E_n(\boldsymbol{\theta})} |J^x_{mn}(\boldsymbol{\theta})|^2\,
\delta\big(E_n(\boldsymbol{\theta})+\omega-E_m(\boldsymbol{\theta})\big),
\label{eq:sigma_regular_without_eta_SM}
\end{equation}
where $\beta=1/T$ is the inverse temperature, $Z(\boldsymbol{\theta})=\sum_n e^{-\beta E_n(\boldsymbol{\theta})}$ is the
partition function of $\hat{H}(\boldsymbol{\theta})$, and $\delta(\cdot)$ is the Dirac delta function. We note that
Eq.~\eqref{eq:sigma_regular_without_eta_SM} is not strictly valid
at $\omega=0$~\cite{liu2026transporttemperature1exact}. For TBCs,
we average the optical conductivity over twists,
\begin{equation}
\sigma_{\mathrm{reg}}^{xx}(\omega,T)
=
\frac{1}{N_{\theta}}
\sum_{\boldsymbol{\theta}}
\sigma_{\mathrm{reg},\boldsymbol{\theta}}^{xx}(\omega,T),
\label{eq:twist_average}
\end{equation}
where the sum runs over the $N_{\theta}=121$ twists of Eqs.~\eqref{eq:actual_twist_grid} and~\eqref{eq:actual_rotated_boundary_twists}.

We note that the real part of the full optical conductivity contains both a
zero-frequency Drude singularity,
$2\pi \mathcal{D}(T)\delta(\omega)$, and the finite-frequency regular part given in
Eq.~\eqref{eq:sigma_regular_without_eta_SM}.
The Drude weight $\mathcal{D}(T)$ includes both diamagnetic and paramagnetic terms
and is typically finite in integrable systems. A nonzero Drude weight
signals ballistic charge transport and hence zero dc resistivity. For the nonintegrable system considered here (see the level statistics in Sec.~\ref{sub:level_statistics}), the Drude weight is
expected to vanish in the thermodynamic limit (although there is no rigorous proof), but finite systems accessible to ED may have a nonzero
Drude weight~\cite{PhysRevLett.132.266502}. The double sum in Eq.~\eqref{eq:sigma_regular_without_eta_SM} is unrestricted and
includes the diagonal terms $m=n$ and the exactly degenerate pairs with $E_m=E_n$.
These terms do not contribute at any nonzero frequency, since the delta function
then requires $E_m-E_n=\omega\neq0$. However, once the delta functions are broadened
as described below [Eq.~\eqref{eq:Lorentzian}], these terms produce a spurious
finite-size Drude-like peak in the optical conductivity near $\omega=0$, which is
visible in Fig.~\ref{fig:details_extrapolation_TBC_window_fit}.
To reduce its influence, we estimate the dc limit of the regular part
by extrapolating finite-frequency data outside this narrow peak
(Sec.~\ref{sec:details_extrapolation_to_dc_limit}).

The energy spectrum of the ED-accessible finite-size systems is discrete,
so the delta function in Eq.~\eqref{eq:sigma_regular_without_eta_SM} must be
broadened to obtain a smooth optical conductivity. A commonly used scheme is Lorentzian broadening~\cite{PhysRevB.105.L201108},
\begin{equation}
\delta_{\eta}^{(L)}(E_n+\omega-E_m)=
\frac{1}{\pi}\frac{\eta}{\eta^2+(E_n+\omega-E_m)^2},
\label{eq:Lorentzian}
\end{equation}
which reduces to the delta function as the broadening parameter $\eta$ is sent to zero,
$\delta(E_n+\omega-E_m)=\lim_{\eta\to0}\delta^{(L)}_\eta(E_n+\omega-E_m)$. Gaussian broadening,
$\delta_{\eta}^{(G)}(E_n+\omega-E_m)
=\frac{1}{\eta\sqrt{\pi}}
\exp\!\left[-\left(\frac{E_n+\omega-E_m}{\eta}\right)^2\right]$,
is also used in the literature~\cite{fratini2026strangemetaltransportcoupling}. We note that the precise values of the optical conductivity obtained with different broadening schemes ($\delta^{(L)}_{\eta}$ and $\delta^{(G)}_{\eta}$), or with different $\eta$ within the same scheme, may differ.
All transport results in this work, including the broadening tests in
Secs.~\ref{sec:spin_sector_dependence_of_the_resistivity}
and~\ref{sec:details_extrapolation_to_dc_limit}, are obtained with the
Lorentzian broadening of Eq.~\eqref{eq:Lorentzian}, and we have not checked
whether Gaussian broadening gives the same results. The Lorentzian kernel has
a simple time-domain interpretation,
\begin{align}
\delta_{\eta}^{(L)}(\omega)
=
\frac{1}{2\pi}
\int_{-\infty}^{\infty}
d\tau\,e^{i\omega \tau}e^{-\eta|\tau|},
\label{eq:Lorentzian_time_domain}
\end{align}
i.e., a Lorentzian of width $\eta$ is equivalent to an exponential cutoff in
the time domain. For the isolated infinite-$U$ Hubbard model considered here,
$\eta$ is an artificial broadening that sets the frequency resolution of the
optical conductivity, not an intrinsic relaxation rate of the
Hamiltonian. Phenomenologically, $\eta$ may be interpreted as an effective
damping rate if an additional relaxation mechanism causes the current--current
correlation function to decay approximately as $e^{-\eta|\tau|}$.

The broadened optical conductivity at fixed twist becomes
\begin{align}
\sigma_{\mathrm{reg},\boldsymbol{\theta}}^{xx}(\omega,T,\eta)
&=
\frac{1-e^{-\beta\omega}}{NZ(\boldsymbol{\theta})\omega}
\sum_n e^{-\beta E_n(\boldsymbol{\theta})}
\sum_m
\left|J_{mn}^{x}(\boldsymbol{\theta})\right|^2
\frac{\eta}
{\eta^2+(E_n(\boldsymbol{\theta})+\omega-E_m(\boldsymbol{\theta}))^2},
\label{eq:sigma_regular_with_eta_L}
\end{align}
and the twist-averaged $\sigma^{xx}_{\mathrm{reg}}(\omega,T,\eta)$ follows from Eq.~\eqref{eq:twist_average}.
We are mainly interested in the dc conductivity, whereas
Eq.~\eqref{eq:sigma_regular_without_eta_SM} applies only at finite frequency.
Formally, the contribution of the regular part to the dc conductivity is
obtained by taking the thermodynamic limit, followed
by the broadening and frequency limits, in the order
\begin{equation}
\sigma^{xx}_{\mathrm{dc}}(T)
=\lim_{\omega\to0}\lim_{\eta\to0}\lim_{N\to\infty}
\sigma^{xx}_{\mathrm{reg}}(\omega,T,\eta).
\label{eq:order_of_limits}
\end{equation}
In ED calculations, however, neither the $N\to\infty$ nor the $\eta\to0$ limit
can be taken literally. The accessible system sizes are severely restricted,
and at fixed $N$, a broadening $\eta$ much smaller than the many-body level
spacing resolves individual transitions, so that the conductivity is sharply
peaked at the transition frequencies and nearly vanishes between them.
Moreover, the limits in Eq.~\eqref{eq:order_of_limits} do not commute: for the
one-dimensional infinite-$U$ Hubbard model, where
$\sigma^{xx}_{\mathrm{reg}}(\omega,T,\eta)$ is known in closed form,
interchanging the order of limits leads to qualitatively different
results~\cite{liu2026transporttemperature1exact}. An explicit example for the $\eta\to0$ and $\omega\to0$ limits is given in Sec.~\ref{sub:analytical_optical_conductivity_with_single_hole_in_fully_polarized_background}.

In our ED calculations, we consider the $N=16$ and $N=18$ square
clusters introduced in
Sec.~\ref{sub:hamiltonan_of_infinite_u_hubbard_model}, each containing
a single doped hole, i.e., at hole densities $1/16$ and $1/18$, respectively. These two clusters do not form a fixed-density
sequence: when the number of doped holes is fixed at one, the hole
density $1/N$ vanishes as $N\to\infty$, and the conductivity
is then expected to vanish in the thermodynamic limit. To extract the dc
conductivity of these finite clusters, we proceed as follows.
\begin{enumerate}
\item \textbf{Choice of the broadening.} At fixed system size $N$,
temperature $T$, and frequency $\omega$, we compute the optical conductivity with a
representative value of $\eta$ (see Secs.~\ref{sec:spin_sector_dependence_of_the_resistivity}
and~\ref{sec:details_extrapolation_to_dc_limit} for more discussion).

\item \textbf{Extraction of the dc limit.} At fixed $N$, $T$, and $\eta$, we
fit $\sigma^{xx}_{\mathrm{reg}}(\omega,T,\eta)$ over a finite-frequency window
$[\omega_{\min},\omega_{\max}]$ with $\omega_{\min}>0$ (see Sec.~\ref{sec:details_extrapolation_to_dc_limit} for the choice of $\omega_{\min}$ and $\omega_{\max}$) to a smooth function of
$\omega$, and define the dc conductivity as the extrapolated value of the fit
at $\omega=0$,
\begin{equation}
\sigma_{\mathrm{dc}}^{xx}(T)
=
\lim_{\omega\to 0}
\sigma_{\mathrm{fit}}^{xx}(\omega,T,\eta).
\label{eq:DC_strategy2}
\end{equation}
Excluding frequencies below $\omega_{\min}$ from the fit reduces the influence
of the spurious finite-size Drude-like peak; see
Sec.~\ref{sec:details_extrapolation_to_dc_limit} for further discussion.

\item \textbf{Resistivity.} Finally, the dc resistivity is obtained by
inversion,
\begin{align}
\rho_x(T)=\frac{1}{\sigma^{xx}_{\mathrm{dc}}(T)} .
\label{eq:rho_from_sigma}
\end{align}
For TBCs, the twist average of Eq.~\eqref{eq:twist_average} is performed
before the fit in Eq.~\eqref{eq:DC_strategy2}. Thus, we first obtain the twist-averaged dc conductivity, which corresponds to averaging the induced current under the same applied electric field, and then invert it to obtain the dc resistivity.
\end{enumerate}

\subsection{Analytical optical conductivity of a single hole in a fully polarized background}
\label{sub:analytical_optical_conductivity_with_single_hole_in_fully_polarized_background}

In the calculation of the energy-resolved level statistics of the infinite-$U$ Hubbard model (see Sec.~\ref{sub:level_statistics}), we have included contributions from all allowed $S^{z}$ sectors: with a single doped hole, $S^{z}= - \frac{N-1}{2}, -\frac{N-3}{2}, \cdots, \frac{N-3}{2}, \frac{N-1}{2}$. In the numerical calculations of charge transport in Secs.~\ref{sec:details_extrapolation_to_dc_limit} and~\ref{sub:temperature_dependence_dc_resistivity}, we also include contributions from all allowed $S^{z}$ sectors with a single doped hole, where the sectors with $S^{z}<0$ are accounted for through the $S^{z}\to-S^{z}$ symmetry (see the beginning of Sec.~\ref{sec:numerical_results}). Here, we instead consider the simplest case with a single doped hole, the fully polarized case with $S^{z} = \pm \frac{N-1}{2}$, which provides an exactly solvable illustration of the Kubo formula in Sec.~\ref{sec:Kubo_formula_and_optical_conductivity}. In these $(N_h, S^{z})=(1, \pm \frac{N-1}{2})$ sectors, all eigenstates of the infinite-$U$ Hubbard model can be obtained analytically.

We consider the sector with $(N_h, S^{z})=(1, \frac{N-1}{2})$, whose Hilbert space dimension is $N$: the position of the hole in the fully polarized background has $N$ choices. The generalization to the sector with $(N_h, S^{z})=(1, -\frac{N-1}{2})$ is straightforward. Let $\ket{\mathrm{FM}}=\prod_{\mathbf{r}}\tilde{c}^{\dagger}_{\mathbf{r},\uparrow}\ket{0}$ denote the fully polarized state at half filling. We use $\ket{x,y}\equiv\tilde{c}_{(x,y),\uparrow}\ket{\mathrm{FM}}$ to denote the basis state where the hole is located at site $(x,y)$ and all other sites are each occupied by a spin-up fermion. Since the PBC Hamiltonian [Eq.~\eqref{eq:H2D_TBCs} with $\boldsymbol{\theta}=\mathbf{0}$] respects the discrete translation symmetry along both the $x$ and $y$ directions, we define the following momentum eigenstates $\ket{k_x, k_y} =  \frac{1}{\sqrt{N}} \sum_{x,y} e^{ -i (k_x x + k_y y) } \ket{x,y}$, where $(k_x,k_y)$ runs over the allowed momenta of the cluster [see Eqs.~\eqref{eq:allowed_momenta_rotated} and~\eqref{eq:momentum_list_rotated} for $N=18$]. The states $\ket{k_x, k_y}$ are eigenstates of the PBC Hamiltonian with energy $2t \cos(k_x) + 2t \cos(k_y)$ and of the current operator in Eq.~\eqref{eq:2DxCurrent} with eigenvalue $2t\sin(k_x)$. The ground state is thus given by $\ket{\pi, \pi}$ with ground-state energy $-4t$ and is also an eigenstate of the current operator with zero eigenvalue. We note that, in the full single-hole Hilbert space, the ground-state degeneracy is $N$ due to the $SU(2)$ symmetry, and all $N$ degenerate ground states can be obtained by applying $(S^{-})^{i}$ to $\ket{\pi, \pi}$ with $i = 0, 1, \cdots, N-1$, where $S^{-}$ is the total spin-lowering operator.

Moreover, in this $(N_{h}, S^z)=(1, \frac{N-1}{2})$ sector, the regular part of the optical conductivity shown in Eq.~\eqref{eq:sigma_regular_with_eta_L} (with $\boldsymbol{\theta} = \boldsymbol{0}$, and with $Z$ and all sums restricted to this sector) has an analytical expression (set $t=1$):
\begin{align}
\sigma^{xx}_{\mathrm{reg}}(\omega, T, \eta) &= \frac{1-e^{-\beta \omega}}{NZ \omega} \sum_{m,n} e^{-\beta E_n} \vert J_{mn}^x \vert^2 \frac{\eta}{\eta^2 + (E_n + \omega - E_m)^2} \nonumber \\
&= \frac{1-e^{-\beta \omega}}{N \omega} \frac{ \sum_{k_x, k_y} e^{- \beta (2 (\cos(k_x) + \cos(k_y)))} (2 \sin(k_x))^2}{\sum_{k_x, k_y} e^{-\beta (2 (\cos(k_x) + \cos(k_y)))}} \frac{\eta}{\eta^2 + \omega^2} \nonumber \\
& = \frac{1-e^{-\beta \omega}}{N \omega} \frac{\eta}{\eta^2 + \omega^2} \frac{ \sum_{k_x, k_y} e^{- \beta (2 (\cos(k_x) + \cos(k_y)))} (2 \sin(k_x))^2}{\sum_{k_x, k_y} e^{-\beta (2 (\cos(k_x) + \cos(k_y)))}},
\label{eq:sigmaregular_polarized}
\end{align}
where in the second equality we have used $\bra{k_x^{\prime}, k_y^{\prime}} \hat{J}^x \ket{k_x, k_y} = 2 \sin(k_x) \delta_{k_x, k_x^{\prime}} \delta_{k_y, k_y^{\prime}}$. The $\omega$ and $\eta$ dependence thus enters only through the overall factor $\frac{1-e^{-\beta\omega}}{N \omega} \frac{\eta}{\eta^2 + \omega^2}$ in the last line of Eq.~\eqref{eq:sigmaregular_polarized}. This example also shows why the $\eta\to0$ and $\omega\to0$ limits in Eq.~\eqref{eq:order_of_limits} do not commute. At fixed $N$, if we first take the $\eta \to 0$ limit at finite frequency $\omega > 0$ and then take the $\omega \to 0$ limit, we have $\lim_{\omega \to 0} \lim_{\eta \to 0} \frac{1-e^{-\beta \omega}}{N \omega} \frac{\eta}{\eta^2 + \omega^2} = 0$, which is consistent with the fact that in this $(N_{h}, S^{z})= (1, \frac{N-1}{2})$ sector, all eigenstates are common eigenstates of $\hat{H}$ and $\hat{J}^{x}$, so that there is no regular-part contribution. In contrast, if we take the limits in the reverse order, we have $\lim_{\eta \to 0} \lim_{\omega \to 0} \frac{1-e^{-\beta \omega}}{N \omega} \frac{\eta}{\eta^2 + \omega^2} = \lim_{\eta \to 0} \frac{\beta}{N \eta} = \infty$, which reflects the zero-frequency Drude weight of this sector (see also Sec.~\ref{sec:spin_sector_dependence_of_the_resistivity} and Ref.~\cite{liu2026transporttemperature1exact}). Moreover, 
in the large-$N$ limit, Eq.~\eqref{eq:sigmaregular_polarized} simplifies to 
\begin{align}
\sigma^{xx}_{\mathrm{reg}}(\omega, T, \eta) \approx \frac{1-e^{-\beta \omega}}{N \beta \omega} \frac{ 2 I_1(2\beta) }{I_0(2\beta)} \frac{\eta}{\eta^2 + \omega^2},
\label{eq:sigmaregular_polarized_largeN}
\end{align}
where $I_n(\cdot)$ denotes the modified Bessel function of the first kind of order $n$ (see Ref.~\cite{liu2026transporttemperature1exact}).

\subsection{\texorpdfstring{Energy-resolved current response and the criterion for strictly linear-in-$T$ resistivity}{Energy-resolved current response and the criterion for strictly linear-in-T resistivity}}
\label{sec:definition_of_f_function}

Before presenting the numerical charge-transport results for the infinite-$U$ Hubbard model in
Secs.~\ref{sec:numerical_results} and~\ref{sec:origin_of_crossover_temperature},
we use the Kubo formula
[Eq.~\eqref{eq:sigma_regular_without_eta_SM}] to determine when the
resistivity is strictly linear in temperature. We call the resistivity strictly
linear in $T$ if $\rho_x(T)=\alpha T$ holds at all $T>0$ with a finite,
positive, temperature-independent coefficient $\alpha$; in particular, the
residual resistivity $\rho_x(T\to0)$ vanishes.

Because the criterion is not specific to the infinite-$U$ Hubbard model, we
formulate it for an arbitrary Hamiltonian $\hat{H}$ with current operator $\hat{J}^{x}$, eigenstates $\ket{n}$, eigenenergies $E_{n}$, and current matrix elements $J^{x}_{mn} = \bra{m} \hat{J}^{x} \ket{n}$. It is useful to introduce the energy-resolved current-response function
$f_x$~\cite{PhysRevB.105.L201108}, which collects the current matrix elements connecting a given eigenstate to the rest
of the spectrum. For the Lorentzian broadening of Eq.~\eqref{eq:Lorentzian} and a given
eigenstate $\ket{n}$, we define
\begin{align}
f_x(\ket{n},\omega,\eta)
&=\sum_m |J^x_{mn}|^2
\frac{\eta}{\eta^2+(E_n+\omega-E_m)^2}.
\label{eq:fx_state_resolved}
\end{align}
 In terms of $f_x(\ket{n},\omega,\eta)$, the broadened
conductivity [Eq.~\eqref{eq:sigma_regular_with_eta_L}] takes the following form
\begin{equation}
\sigma^{xx}_{\mathrm{reg}}(\omega,T,\eta)
=\frac{1-e^{-\beta\omega}}{N\omega}
\frac{\sum_n e^{-\beta E_n}\,f_x(\ket{n},\omega,\eta)}
{\sum_n e^{-\beta E_n}} .
\label{eq:sigma_from_fx}
\end{equation}
We label the $M$ \emph{distinct} eigenenergies of $\hat{H}$ by $\{\varepsilon_i\}_{i=1}^{M}$, so that
$\varepsilon_i\neq\varepsilon_j$ for $i\neq j$, and denote the degeneracy of
$\varepsilon_i$ by $g_i\ge1$. We
define the energy-resolved current-response function $f_x$ by averaging over the states with
the same eigenenergy,
\begin{equation}
f_x(\varepsilon_i,\omega,\eta)
=\frac{1}{g_i}\sum_{n\,:\, E_n=\varepsilon_i}
f_x(\ket{n},\omega,\eta).
\label{eq:f_function_SM}
\end{equation}

The optical conductivity [Eq.~\eqref{eq:sigma_from_fx}] then
reads
\begin{align}
\sigma^{xx}_{\mathrm{reg}}(\omega,T,\eta)
=\frac{1-e^{-\beta\omega}}{N\omega}
\frac{\sum_i g_i\,e^{-\beta \varepsilon_i}\,f_x(\varepsilon_i,\omega,\eta)}
{\sum_i g_i\,e^{-\beta \varepsilon_i}} .
\label{eq:sigma_from_fx_avg}
\end{align}
The prefactor of Eq.~\eqref{eq:sigma_from_fx_avg} tends to $\beta/N$ as $\omega\to0$, and we
define $f_x(\varepsilon_i,0,\eta)\equiv\lim_{\omega\to0}f_x(\varepsilon_i,\omega,\eta)$,
which is independent of temperature. The corresponding dc
conductivity is
\begin{equation}
\sigma^{xx}_{\mathrm{reg}}(0,T,\eta)
=\frac{\beta}{N}
\frac{\sum_{i} g_i\,e^{-\beta \varepsilon_i}\,f_x(\varepsilon_i,0,\eta)}
{\sum_{i} g_i\,e^{-\beta \varepsilon_i}},
\label{eq:sigma_dc_level_resolved}
\end{equation}
and the dc resistivity is
\begin{equation}
\rho_x(T)
=N T\,
\frac{\sum_{i} g_i\,e^{-\beta \varepsilon_i}}
{\sum_{i} g_i\,e^{-\beta \varepsilon_i}\,f_x(\varepsilon_i,0,\eta)}.
\label{eq:rho_level_resolved}
\end{equation}

\begin{theorem}
\label{prop:fx_linear_T}
For a given Hamiltonian and constant
broadening $\eta$, the resistivity
of Eq.~\eqref{eq:rho_level_resolved} is strictly linear in temperature,
$\rho_x(T)=\alpha T$ for all $T>0$ with a temperature-independent
$0<\alpha<\infty$, if and only if
$f_x(\varepsilon_i,0,\eta)=f_0$ for all $i$, with $0<f_0<\infty$. In that case, $\alpha=N/f_0$.
\end{theorem}

\begin{proof}
\textbf{Sufficiency.} If $f_x(\varepsilon_i,0,\eta)=f_0$ is independent of $i$, the two sums in
Eq.~\eqref{eq:rho_level_resolved} cancel and
\begin{equation}
\rho_x(T)=\frac{N}{f_0}\,T ,
\label{eq:rho_constant_fx}
\end{equation}
which is strictly linear in $T$ with coefficient $\alpha=N/f_0$.

\textbf{Necessity.} As an alternative to the original derivation in
Ref.~\cite{PhysRevB.105.L201108}, we give a proof based on a Vandermonde matrix
that accounts for degeneracies explicitly. Assume $\rho_x(T)=\alpha T$ for all $T>0$ and define
$f_0\equiv N/\alpha$. Inserting this into Eq.~\eqref{eq:rho_level_resolved} and
rearranging gives
\begin{equation}
G(\beta)\equiv\sum_{i=1}^{M} u_i\,e^{-\beta\varepsilon_i}=0
\quad\text{for all }\beta>0,
\qquad
u_i\equiv g_i\left(\frac{f_x(\varepsilon_i,0,\eta)}{f_0}-1\right).
\label{eq:G_tau}
\end{equation}
Since $G(\beta)$ is a finite linear combination of exponential
functions, it is entire~\cite[Sec.~4.2(iii)]{NIST:DLMF}.
Because $G(\beta)=0$ for every $\beta>0$, all its derivatives
vanish on this interval and, by continuity, at $\beta=0$. Evaluating $G$ and its first $M-1$ derivatives at $\beta=0$
yields
\begin{equation}
G^{(k)}(0)=\sum_{i=1}^{M}(-\varepsilon_i)^{k}\,u_i=0,
\qquad k=0,1,\ldots,M-1,
\label{eq:G_derivatives}
\end{equation}
which is the linear system $V\mathbf{u}=\mathbf{0}$ with
$\mathbf{u}=(u_1,u_2,\ldots,u_M)^{\mathsf{T}}$ and
\begin{equation}
V=
\begin{pmatrix}
1 & 1 & \cdots & 1\\
-\varepsilon_1 & -\varepsilon_2 & \cdots & -\varepsilon_M\\
(-\varepsilon_1)^{2} & (-\varepsilon_2)^{2} & \cdots & (-\varepsilon_M)^{2}\\
\vdots & \vdots & \ddots & \vdots\\
(-\varepsilon_1)^{M-1} & (-\varepsilon_2)^{M-1} & \cdots &
(-\varepsilon_M)^{M-1}
\end{pmatrix}.
\label{eq:vandermonde_matrix}
\end{equation}
The matrix $V$ is a Vandermonde matrix in the variables $-\varepsilon_i$, so that
\begin{equation}
\det V=\prod_{1\le i<j\le M}\left(\varepsilon_i-\varepsilon_j\right)\neq0,
\label{eq:vandermonde_det}
\end{equation}
because the $\varepsilon_i$ are distinct by construction. Hence $V$ is
invertible and $\mathbf{u}=\mathbf{0}$. Since $g_i\ge1$, this forces
\begin{equation}
\frac{f_x(\varepsilon_i,0,\eta)}{f_0}-1=0
\quad\Longrightarrow\quad
f_x(\varepsilon_i,0,\eta)=f_0
\quad\text{for all }i ,
\label{eq:fx_constant_result}
\end{equation}
i.e., $f_x(\varepsilon_i,0,\eta)$ is energy independent.
\end{proof}

We note that the criterion established above applies at a fixed twist.
For TBCs, we first calculate the conductivity at each twist, normalized by its own partition function, and then average over twists,
as in Eq.~\eqref{eq:twist_average}. This procedure generally yields
a different result from combining the eigenstates of all twists into a single partition function before calculating the conductivity. In addition, temperature-dependent
variations in the conductivities at different twists can cancel upon
averaging, so the necessity part of Theorem~\ref{prop:fx_linear_T} does not
automatically extend to the twist-averaged resistivity. We therefore use the
twist-averaged energy-resolved $f_x$ shown in Fig.~\ref{fig:fx_dos_N16_N18_TBC}
(Sec.~\ref{sub:energy_resolved_f_x_function}) as a diagnostic of the spectral structure, rather than as a necessary and
sufficient criterion. In addition, Theorem~\ref{prop:fx_linear_T} concerns only strict linearity,
$\rho_x(T)=\alpha T$ for all $T>0$. It does not imply that a constant dc $f_x$ is
necessary for the resistivity to be approximately linear in $T$ over a finite
temperature range.

Moreover, strict energy independence is a strong requirement, and in practice only
approximately linear-in-$T$ resistivity is observed. The theorem thus
serves only as a qualitative guide: a flatter $f_x(\varepsilon)$ is associated with more nearly linear-in-$T$
behavior. We do not have a quantitative criterion that relates the flatness
of $f_x$ to a given tolerance on the linearity of the resistivity, such as
$5\%$ or $10\%$. For the ED resistivity, the linear fits, their $R^2$ values,
and the maximum relative residuals are given in
Sec.~\ref{sub:temperature_dependence_dc_resistivity}.
As shown in Sec.~\ref{sub:energy_resolved_f_x_function}, for the
infinite-$U$ Hubbard model the dc $f_x$ extrapolated from the ED data is nearly
constant in the interior of the spectrum but varies strongly in edge regions
of width $\varepsilon_{f_x}$. This edge scale $\varepsilon_{f_x}$ is related to the crossover
temperature $T_{\mathrm{cr}}$, which governs how the resistivity departs
from its high-temperature linear form as the temperature is lowered; this relation is discussed in detail in Sec.~\ref{sub:energy_resolved_f_x_function}.

\section{Exact diagonalization results for the dc resistivity}
\label{sec:numerical_results}

Using the formalism of Sec.~\ref{sec:model_and_transport_formalism}, we
present in this section the ED results for the resistivity of the
infinite-$U$ Hubbard model with a single doped hole. In
Sec.~\ref{sec:spin_sector_dependence_of_the_resistivity}, we use the $4\times4$
cluster with PBCs to show the dependence of the optical conductivity on the
broadening parameter $\eta$ and on the spin sector.
In Sec.~\ref{sec:details_extrapolation_to_dc_limit}, we describe the
finite-frequency extrapolation used to extract the dc limit and test its
robustness. In Sec.~\ref{sub:temperature_dependence_dc_resistivity}, we present
the temperature dependence of the extrapolated dc resistivity and identify two approximately linear-in-$T$ resistivity regimes.

We first specify the spin sectors retained in these calculations. With a single
doped hole, there are $N$ distinct $S^z$ sectors,
$S^{z} = -\frac{N-1}{2}, -\frac{N-3}{2}, \ldots, \frac{N-3}{2}, \frac{N-1}{2}$,
corresponding to $N_{\downarrow}=N-1,N-2,\ldots,0$ spin-down fermions. In the ED
calculations of the optical conductivity presented in
Secs.~\ref{sec:details_extrapolation_to_dc_limit}
and~\ref{sub:temperature_dependence_dc_resistivity}, we exploit the symmetry
between the sectors with positive and negative $S^z$: the sectors with
$S^{z} = \frac{1}{2}, \frac{3}{2}, \ldots, \frac{N-1}{2}$ and those with
$S^{z} = -\frac{1}{2}, -\frac{3}{2}, \ldots, -\frac{N-1}{2}$ contribute equally to
the partition function, and they also contribute equally to the current response
in Eq.~\eqref{eq:sigma_regular_with_eta_L}, i.e., $\sum_n e^{-\beta E_n(\boldsymbol{\theta})} \sum_m
\left|J_{mn}^{x}(\boldsymbol{\theta})\right|^2
\frac{\eta}{\eta^2+(E_n(\boldsymbol{\theta})+\omega-E_m(\boldsymbol{\theta}))^2}$.
For the even-$N$ clusters considered here ($N=16,18$), we therefore retain only the
sectors with positive $S^{z} = \frac{1}{2}, \frac{3}{2}, \ldots, \frac{N-1}{2}$ in
Eq.~\eqref{eq:sigma_regular_with_eta_L} to reduce the computational cost; the
resulting factor of two cancels between the numerator and the denominator.
Specifically, at each fixed twist $\boldsymbol{\theta}$, both the partition function
in the denominator and the double sum over $m$ and $n$ in
Eq.~\eqref{eq:sigma_regular_with_eta_L} are evaluated over all eigenstates of the
Hamiltonian in Eq.~\eqref{eq:H2D_TBCs} at that twist that lie in the retained
positive-$S^z$ sectors.

We note that the same restriction to the positive-$S^z$ sectors is also used to
compute the energy-resolved current-response function $f_x$ in
Sec.~\ref{sub:energy_resolved_f_x_function}. Owing to the symmetry between the
positive- and negative-$S^z$ sectors, the $f_x$ obtained by retaining only the
positive-$S^z$ sectors is the same as that obtained by retaining both. The density of
states shown in Sec.~\ref{sub:density_of_states_and_specific_heat} counts only the
states in the retained positive-$S^z$ sectors, so the full density of states is
twice the displayed value.

\subsection{Dependence on the broadening and the spin sector under PBCs}
\label{sec:spin_sector_dependence_of_the_resistivity}

In this subsection, we use the
conventional $4\times4$ square cluster (see Fig.~\ref{fig:Rotated_Lattice_N18}(a)) with PBCs
($\boldsymbol{\theta}=\mathbf{0}$) and Lorentzian broadening
to show how the optical conductivity depends on
(i) the $S^z$ sector and (ii) the broadening parameter $\eta$.
In these $S^z$-resolved calculations, Eq.~\eqref{eq:sigma_regular_with_eta_L}
is evaluated within a given $S^z$ sector, i.e., at fixed $N_{\downarrow}$, since $S^{z} = \frac{N-1-2N_{\downarrow}}{2}$,
for both the partition function and the sum over eigenstates.

\begin{figure}[t]
\centering
\includegraphics[width=7in]{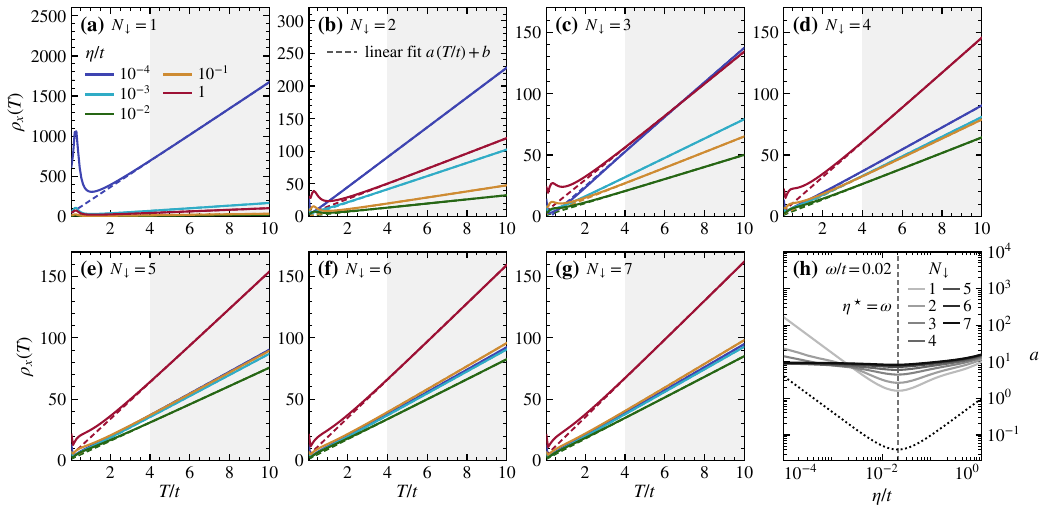}
\caption{Dependence of the optical resistivity on $\eta$ and on the $S^z$ sector. Optical resistivity $\rho_x(T)\equiv\rho_x(\omega,T,\eta)$ at $\omega/t=0.02$ for
the $4\times4$ cluster with PBCs and a single doped hole, over $0.05\leq T/t\leq10$. We computed $\rho_x$ for $39$ Lorentzian broadenings over
$10^{-4}\leq\eta/t\leq1$ [$\eta/t=(1,2,3,5,7)\times10^{-4}$,
$(1,1.3,1.5,2,2.5,3,4,5,6,7,8)\times10^{-3}$,
$(1,1.2,1.5,1.8,2,2.5,3,4,5,6,7,8)\times10^{-2}$,
$(1,1.2,1.5,1.8,2,2.5,3,4,5,7)\times10^{-1}$, and $1$]; for clarity, panels (a)--(g), corresponding to the $S^z$
sectors with $N_{\downarrow}=1,\ldots,7$, respectively, show only
$\eta/t=10^{-4}$, $10^{-3}$, $10^{-2}$, $10^{-1}$, and $1$. Dashed lines are linear
fits $\rho_x(T)=a\,(T/t)+b$ over $4\leq T/t\leq10$ (shaded region), extended to lower
temperatures. (h) Fitted slope $a$ versus $\eta/t$ for all $39$ broadenings for each $S^z$ sector with given $N_{\downarrow}$. The
dotted curve is proportional to $[(\eta/t)^2+(\omega/t)^2]/(\eta/t)$, the $\eta$
dependence of the optical resistivity in the fully polarized sector
[Eq.~\eqref{eq:sigmaregular_polarized}], and the vertical dashed line marks its
minimum at $\eta^{\star}=\omega$.}
\label{fig:fixedL4_omega_0p02_Ndown_panels}
\end{figure}

In the fully polarized sector $S^{z}=\frac{N-1}{2}$, i.e., $N_{\downarrow}=0$, all
eigenstates can be obtained analytically (see
Sec.~\ref{sub:analytical_optical_conductivity_with_single_hole_in_fully_polarized_background}), and we have obtained the analytical expression for the optical conductivity in this $S^z = \frac{N-1}{2}$ sector [Eqs.~\eqref{eq:sigmaregular_polarized} and~\eqref{eq:sigmaregular_polarized_largeN}].
Equations~\eqref{eq:sigmaregular_polarized} and~\eqref{eq:sigmaregular_polarized_largeN}
show that, in this $S^z=\frac{N-1}{2}$ sector, the $\omega$ and $\eta$ dependence of the broadened optical conductivity enters only through an overall factor $\frac{1-e^{-\beta \omega}}{N \omega} \frac{\eta}{\eta^2 + \omega^2}$, and thus the broadened optical conductivity depends strongly on the choice of $\eta$. This strong $\eta$ dependence persists in $S^z$ sectors with a
small number of spin-down fermions, such as $N_{\downarrow}=1$ and $2$, and only the $S^z$ sectors with comparable numbers of spin-up and
spin-down fermions show a finite range of $\eta$ over which the optical conductivity
is relatively insensitive to the broadening. See the detailed discussion below and Fig.~\ref{fig:fixedL4_omega_0p02_Ndown_panels}.

We also recall the two limiting regimes of $\eta$. Here the many-body level spacing of the Hamiltonian
is $\delta_{\mathrm{mb}}(E)\sim e^{-S(E)}$, where $S(E)=\ln N_E$ is the microcanonical
entropy and $N_E$ is the number of many-body states in a narrow energy window around
$E$. As noted in Sec.~\ref{sec:Kubo_formula_and_optical_conductivity}, an $\eta$
much smaller than the many-body level spacing resolves individual transitions, so
that the optical conductivity consists of sharp peaks rather than a smooth response.
An overly large $\eta$, on the other hand, washes out the spectral structure.
These two limits give only a qualitative guide. The level spacing depends on
the energy and on the symmetry sector, whereas we use a single,
energy-independent $\eta$, and we do not match $\eta$ quantitatively to the
level spacing. Instead, we choose $\eta$ from the sensitivity of the optical
conductivity to the broadening, using
Fig.~\ref{fig:fixedL4_omega_0p02_Ndown_panels} and
Table~\ref{tab:finite_frequency_mean_slopes} for the $4\times4$ cluster with
PBCs, and the tests in Sec.~\ref{sec:details_extrapolation_to_dc_limit} for
the twist-averaged conductivity. Even where the $\eta$ dependence is weak,
conductivities obtained with different $\eta$ differ slightly.

We denote the optical resistivity at finite frequency by
$\rho_x(\omega,T,\eta)\equiv
[\sigma_{\mathrm{reg}}^{xx}(\omega,T,\eta)]^{-1}$. Figure~\ref{fig:fixedL4_omega_0p02_Ndown_panels}
shows the optical resistivity $\rho_x(\omega,T,\eta)$ at $\omega/t=0.02$ in the $S^z$ sectors with
$N_{\downarrow}=1,\ldots,7$, shown in panels (a)--(g) of Fig.~\ref{fig:fixedL4_omega_0p02_Ndown_panels}, respectively. We make two observations. First,
in the fully polarized sector with $N_{\downarrow}=0$, the analytical expressions for the optical conductivity in this $S^z$ sector [Eqs.~\eqref{eq:sigmaregular_polarized} and~\eqref{eq:sigmaregular_polarized_largeN}] show that, at fixed system size and fixed nonzero $\omega$, the overall factor $\frac{1-e^{-\beta \omega}}{N \omega} \frac{\eta}{\eta^2+\omega^2}$ in Eqs.~\eqref{eq:sigmaregular_polarized} and~\eqref{eq:sigmaregular_polarized_largeN} vanishes as
$\eta\to0$. Consequently, as $\eta$ decreases,
the optical conductivity vanishes and the optical resistivity
diverges. Although we do not have analytical expressions for the optical conductivity beyond the fully polarized sector, we observe that in $S^z$ sectors with only a few spin-down fermions, the optical conductivity and optical resistivity behave similarly to those in the fully polarized sector:
for $N_{\downarrow}=1$ and $2$ [Figs.~\ref{fig:fixedL4_omega_0p02_Ndown_panels}(a)
and~\ref{fig:fixedL4_omega_0p02_Ndown_panels}(b)], the optical resistivity
depends strongly on $\eta$ and increases as $\eta$ is reduced. Second, as
$N_{\downarrow}$ increases toward the minimally polarized $S^z=1/2$ sector,
the $\eta$ dependence becomes substantially weaker, and a finite interval
emerges over which the optical resistivity depends only weakly on $\eta$. To quantify the $\eta$ dependence in each $S^z$ sector, we fit the optical resistivity $\rho_x(\omega=0.02t,T,\eta)$ at given $\eta$ and $N_{\downarrow}$ linearly over the temperature range $T/t \in [4,10]$:
\begin{align}
\rho_x(\omega=0.02t,T,\eta) = a (T/t) + b,
\label{eq:linearfitting}
\end{align}
where $a$ and $b$ are the fitted slope and intercept, respectively. In panels (a)--(g) of Fig.~\ref{fig:fixedL4_omega_0p02_Ndown_panels}, the solid lines show the numerically computed optical resistivity for given $\eta$ and $N_{\downarrow}$, and the dashed lines show the corresponding linear fits [Eq.~\eqref{eq:linearfitting}].
Figure~\ref{fig:fixedL4_omega_0p02_Ndown_panels}(h) shows the fitted $a$ versus $\eta/t$ for each $S^z$ sector. The dotted curve is proportional to $(\eta^2+\omega^2)/\eta$, the expected $\eta$ dependence in the fully polarized sector with $N_{\downarrow}=0$ [Eqs.~\eqref{eq:sigmaregular_polarized} and~\eqref{eq:sigmaregular_polarized_largeN}]. For $N_{\downarrow}=1$ and $2$, the fitted $a$ as a function of $\eta$ has a shape similar to that for $N_{\downarrow}=0$ (dotted curve). In the minimally polarized $S^z=1/2$ sector with $N_{\downarrow}=7$, $N_{\uparrow}=8$, and $N_h=1$, the fitted $a$ depends only weakly on $\eta$. Over $10^{-4}\leq\eta/t\leq10^{-1}$, $a$ deviates from its mean $\bar{a}$ by at most $8.63\%$ (Table~\ref{tab:finite_frequency_mean_slopes}). We therefore describe the $\eta$ dependence in this sector as weak rather than absent. These results also show that, although the optical conductivity can be obtained analytically in the simplest fully polarized sector, this sector does not capture the strong correlations in the more complicated sectors with comparable numbers of spin-up and spin-down fermions. At $\omega/t=0.04$ and $0.06$, the dependence of the optical resistivity on $\eta$ and $S^z$ is qualitatively the same as at $\omega/t=0.02$. For $N_{\downarrow}=7$, Table~\ref{tab:finite_frequency_mean_slopes} lists $\bar{a}$ and the maximum relative deviation of $a$ from $\bar{a}$ at all three frequencies. The mean slopes at $\omega/t=0.04$ and $0.06$ are larger than that at $\omega/t=0.02$ by $0.39\%$ and $4.36\%$, respectively.

\begin{table}[tb]
\caption{Mean high-temperature slope $\bar{a}$ of the optical resistivity
in the $S^z=1/2$ sector ($N_{\downarrow}=7$) of the $4\times4$ cluster with
PBCs and a single doped hole. At each $\omega$ and $\eta$, the slope $a$ is
obtained from an unweighted least-squares fit to $\rho_x=a(T/t)+b$ over
$4\leq T/t\leq10$. The mean $\bar{a}$ is the arithmetic average of $a$ over
the $29$ broadenings with $10^{-4}\leq\eta/t\leq10^{-1}$ listed in the
caption of Fig.~\ref{fig:fixedL4_omega_0p02_Ndown_panels}. The last column
gives the maximum relative deviation of $a$ from $\bar{a}$ over these
broadenings, in percent.}
\label{tab:finite_frequency_mean_slopes}
\centering
\begin{ruledtabular}
\begin{tabular}{ccc}
$\omega/t$ & $\bar{a}$ & $100\max_{\eta}|a-\bar{a}|/\bar{a}$ \\
0.02 & 8.8744 & 8.63 \\
0.04 & 8.9088 & 9.05 \\
0.06 & 9.2609 & 6.04 \\
\end{tabular}
\end{ruledtabular}
\end{table}

\begin{figure}[t]
\centering
\includegraphics[width=7in]{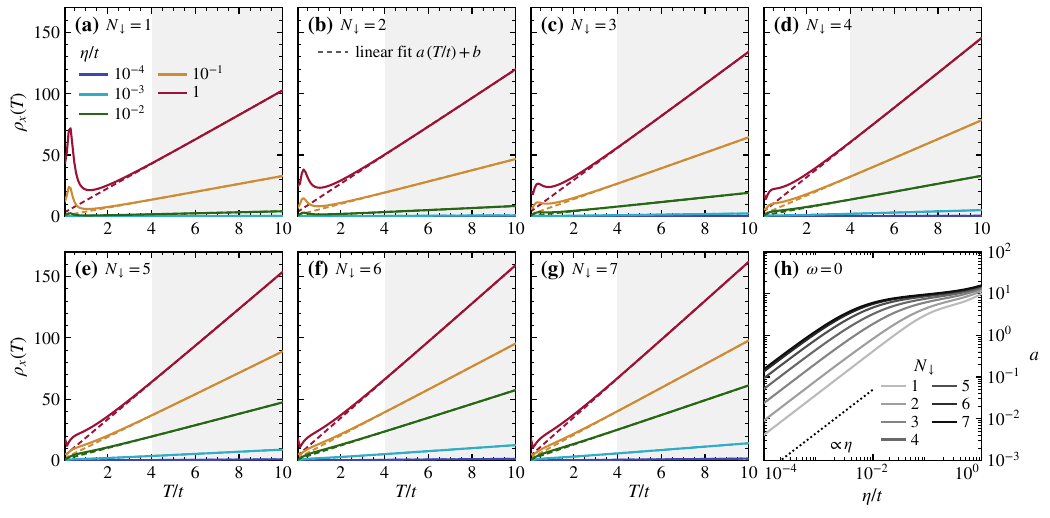}
\caption{Zero-frequency resistivity
$\rho_x(T)\equiv[\sigma^{xx}_{\mathrm{reg}}(0,T,\eta)]^{-1}$ obtained by direct
substitution [Eq.~\eqref{eq:zero_frequency_conductivity}] for the $4\times4$ cluster
with PBCs and a single doped hole, over $0.05\leq T/t\leq10$. In each $S^z$ sector with given $N_{\downarrow}$ [panels (a)--(g)], the resistivity decreases as $\eta$ decreases. As in
Fig.~\ref{fig:fixedL4_omega_0p02_Ndown_panels}, we computed $\rho_x$ for $39$ Lorentzian
broadenings in the range $10^{-4}\leq\eta/t\leq1$; for clarity, panels (a)--(g),
corresponding to $N_{\downarrow}=1,\ldots,7$, respectively, show only
$\eta/t=10^{-4}$, $10^{-3}$, $10^{-2}$, $10^{-1}$, and $1$. Dashed lines are linear
fits $\rho_x(T)=a\,(T/t)+b$ over $4\leq T/t\leq10$ (shaded region), extended to lower
temperatures. (h) Fitted slope $a$ versus $\eta/t$ for all $39$ broadenings. The
dotted line is proportional to $\eta/t$, the $\eta$ dependence of the zero-frequency resistivity in the fully polarized sector
[Eq.~\eqref{eq:sigmaregular_polarized} in the limit $\omega\to0$].}
\label{fig:fixedL4_omega0_Ndown_panels}
\end{figure}

Moreover, as noted in Sec.~\ref{sec:Kubo_formula_and_optical_conductivity},
Eq.~\eqref{eq:sigma_regular_without_eta_SM} cannot be evaluated by simply
substituting $\omega=0$ (see Ref.~\cite{liu2026transporttemperature1exact} for the
detailed derivation). We now illustrate the resulting numerical problem. Taking
$\lim_{\omega\to0}(1-e^{-\beta\omega})/\omega=\beta$ in the prefactor and directly
setting $\omega=0$ in the Lorentzian of Eq.~\eqref{eq:sigma_regular_with_eta_L} gives
\begin{equation}
\sigma^{xx}_{\mathrm{reg}}(0,T,\eta)
= \frac{\beta}{NZ}
\sum_{m,n} e^{-\beta E_n}|J^x_{mn}|^2
\frac{\eta}{\eta^2+(E_n-E_m)^2}.
\label{eq:zero_frequency_conductivity}
\end{equation}
Figure~\ref{fig:fixedL4_omega0_Ndown_panels} shows the corresponding resistivity
$\rho_x(T)=1/\sigma^{xx}_{\mathrm{reg}}(0,T,\eta)$. Panels (a)--(g), for
$N_{\downarrow}=1,\ldots,7$, show $\rho_x(T)$ (solid lines) and its linear fit over
$4\leq T/t\leq10$ (dashed lines), and panel (h) summarizes the fitted slope $a$
versus $\eta/t$. In contrast to the finite-frequency results at $\omega/t=0.02$
(Fig.~\ref{fig:fixedL4_omega_0p02_Ndown_panels}), the resistivity obtained by direct
substitution depends strongly on $\eta$ in every $S^z$ sector, including $S^z=1/2$.
In the fully polarized sector ($N_{\downarrow}=0$),
Eqs.~\eqref{eq:sigmaregular_polarized} and~\eqref{eq:sigmaregular_polarized_largeN}
at $\omega=0$ contain an overall factor $1/\eta$. The resistivity is therefore
proportional to $\eta$ and vanishes as $\eta\to0$, opposite to the finite-frequency
optical resistivity, which diverges in this limit. Although no analytical expression
is available for the other sectors, Figs.~\ref{fig:fixedL4_omega0_Ndown_panels}(a)--(g)
show the same trend: the resistivity decreases as $\eta$ decreases. Accordingly, the
fitted slope $a$ in Fig.~\ref{fig:fixedL4_omega0_Ndown_panels}(h) depends strongly on
$\eta$ in every sector and becomes proportional to $\eta$ at small $\eta$, as in the
fully polarized sector (dotted reference line $\propto\eta$). Direct substitution of
$\omega=0$ therefore does not yield a broadening-independent dc limit, and in
Sec.~\ref{sec:details_extrapolation_to_dc_limit} we instead obtain the dc
conductivity and resistivity by finite-frequency extrapolation.

\subsection{Finite-frequency extrapolation of the twist-averaged conductivity to the dc limit}
\label{sec:details_extrapolation_to_dc_limit}

Guided by the PBC tests in
Sec.~\ref{sec:spin_sector_dependence_of_the_resistivity}, we now turn to the
twist-averaged optical conductivity of the $N=16$ and
$18$ clusters under TBCs, computed by including in Eq.~\eqref{eq:sigma_regular_with_eta_L} all eigenstates in the positive-$S^z$ sectors with $S^z = \frac{1}{2}, \frac{3}{2}, \ldots, \frac{N-1}{2}$, and detail the
finite-frequency extrapolation used to extract the dc conductivity and
resistivity.

For both the $N=16$ and $18$ clusters, we first examine how the
twist-averaged optical conductivity, averaged over the $N_{\theta}=121$ twists according to
Eq.~\eqref{eq:twist_average}, depends on the Lorentzian broadening
$\eta$. We consider the seven frequencies used in the dc
extrapolation in Eq.~\eqref{eq:quadratic_dc_extrapolation}, $\omega/t=0.025,0.030,0.035,0.040,0.045,0.050,0.055$,
at temperatures $0.05 \leq T/t \leq 16$ (the choice of the lowest temperature $T/t=0.05$ is discussed in Sec.~\ref{sub:temperature_dependence_dc_resistivity}).
We test five broadenings,
$\eta/t=0.005$, $0.01$, $0.015$, $0.02$, and $0.1$, and use the twist-averaged optical conductivity $\sigma^{xx}_{\mathrm{reg}}(\omega, T, \eta=0.005t)$ as the reference. At given $\omega$ and $T$, we quantify the change of the optical conductivity for a different $\eta$ by the relative deviation $|\sigma^{xx}_{\mathrm{reg}}(\omega,T,\eta)-\sigma^{xx}_{\mathrm{reg}}(\omega,T,\eta = 0.005t)|
/\sigma^{xx}_{\mathrm{reg}}(\omega,T,\eta=0.005t)$. For $\eta/t=0.1$, the relative deviations at all seven frequencies are below $10 \%$ for $T/t\gtrsim2.23$. For smaller broadenings, the relative deviations at all seven frequencies are below $10 \%$ for $T/t\gtrsim0.1$, $0.15$, and $0.17$ when $\eta/t=0.01$, $0.015$, and $0.02$, respectively. In particular, the optical conductivities with $\eta/t=0.005$ and $\eta/t =0.01$ differ only weakly at all seven frequencies over most of the temperature range. We note that for $0.05 \leq T/t \lesssim 0.1$, the relative changes of the optical conductivity are still higher than $10\%$, so that the optical conductivity is more sensitive to the choice of $\eta$ in the low-temperature regime. One may also use the optical conductivities at given $\omega$ and $T$ within a chosen range of $\eta$ to extrapolate to $\eta=0$, but the result depends on the range of $\eta$ chosen. Instead, we fix $\eta/t=0.005$ as a representative choice for the dc extrapolation [Eq.~\eqref{eq:quadratic_dc_extrapolation}] and for the energy-resolved $f_x$ (Sec.~\ref{sub:energy_resolved_f_x_function}).

\begin{figure}[t]
\centering
\includegraphics[width=7in]{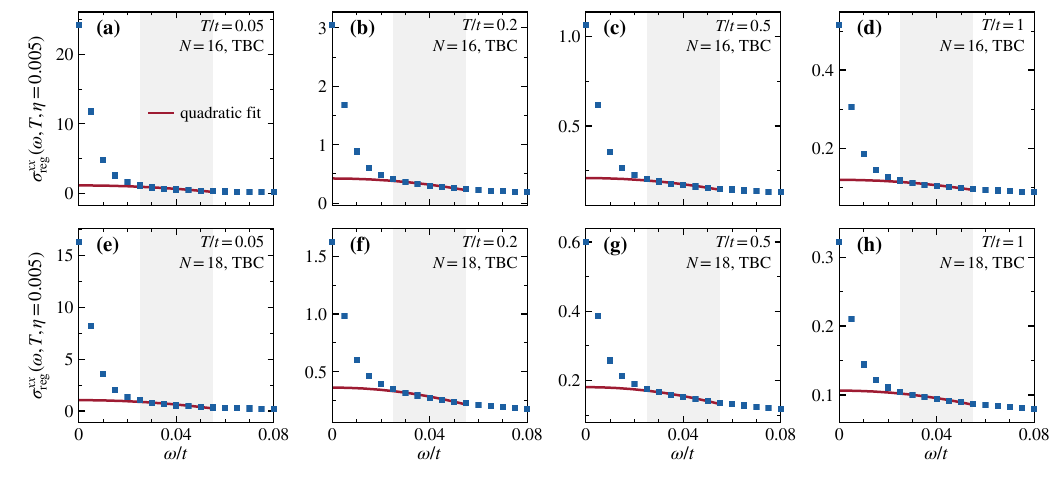}
\caption{Extrapolation of the optical conductivity $\sigma^{xx}_{\mathrm{reg}}(\omega, T, \eta =0.005t)$ to the dc limit under TBCs with Lorentzian
broadening $\eta/t=0.005$. The upper and lower rows show the $N=16$ and
$N=18$ clusters, respectively; from left to right, the columns correspond to
$T/t=0.05$, $0.2$, $0.5$, and $1.0$. Black squares show the twist-averaged
optical conductivity $\sigma_{\mathrm{reg}}^{xx}(\omega,T, \eta =0.005t)$. Red curves
show the quadratic fits of Eq.~\eqref{eq:quadratic_dc_extrapolation}
in the shaded window $0.025\leq\omega/t\leq0.055$. The points at
$\omega=0$, obtained by direct substitution [Eq.~\eqref{eq:zero_frequency_conductivity}], and the points outside $0.025\leq\omega/t\leq0.055$ are not included in the
fits.}
\label{fig:details_extrapolation_TBC_window_fit}
\end{figure}

To obtain the dc conductivity and the dc resistivity, for a given square cluster ($N=16,18$), temperature $T$, and frequency $\omega$, we first average the
optical conductivity over the $N_{\theta}=121$ twists according to
Eq.~\eqref{eq:twist_average} to obtain the optical conductivity $\sigma^{xx}_{\mathrm{reg}}(\omega, T, \eta = 0.005t)$. For a given cluster and temperature $T$, we then fit the
$\sigma_{\mathrm{reg}}^{xx}(\omega,T, \eta = 0.005t)$ at the seven frequencies $\omega/t = 0.025,0.030,0.035,0.040,0.045,0.050,0.055$, i.e., over $\omega_{\min}/t\leq\omega/t\leq\omega_{\max}/t$ with
$\omega_{\min}/t=0.025$ and $\omega_{\max}/t=0.055$, to the quadratic form
\begin{equation}
 \sigma_{\mathrm{fit}}^{xx}(\omega,T, \eta = 0.005t)
 =\sigma_{\mathrm{dc}}^{xx}(T)+c_2(T)\,(\omega/t)^2,
 \label{eq:quadratic_dc_extrapolation}
\end{equation}
where the fitted intercept $\sigma_{\mathrm{dc}}^{xx}(T)$ is the extrapolated dc conductivity. Figure~\ref{fig:details_extrapolation_TBC_window_fit} illustrates this extrapolation
procedure for both clusters ($N=16,18$) at four representative temperatures, $T/t=0.05$, $0.2$, $0.5$, and $1.0$, where the gray region marks the frequencies used in the extrapolation. We note that the data points of the conductivity at $\omega = 0$
are obtained from the
direct-substitution expression in
Eq.~\eqref{eq:zero_frequency_conductivity}. We also note that the finite-broadening expression [Eq.~\eqref{eq:sigma_regular_with_eta_L}] used here
need not be even in $\omega$; we therefore also test a linear fit over the same frequency window. For all
temperatures $T/t\geq0.05$, the resulting dc resistivity differs from that obtained from the
quadratic fit in Eq.~\eqref{eq:quadratic_dc_extrapolation} by at most about $3\%$.

In the extrapolation of the optical conductivity, a finite-frequency window is essential. As discussed in
Sec.~\ref{sec:spin_sector_dependence_of_the_resistivity}, under direct substitution at
$\omega=0$, both the conductivity and the resistivity depend strongly on $\eta$. At small
nonzero $\omega$, the optical conductivity exhibits the spurious finite-size
Drude-like peak discussed in Sec.~\ref{sec:Kubo_formula_and_optical_conductivity}, which is expected to vanish in the thermodynamic limit, where the $\eta \to 0$ limit can be taken literally. In finite-size systems with a given finite $\eta$, however, we need to choose $\omega_{\min}$ in the extrapolation to mitigate the influence of this spurious finite-size Drude-like peak. A residual contribution from the broadened tail may remain within the fitting window. At much larger frequencies, the optical conductivity is no longer
representative of the low-frequency behavior relevant to the extrapolation of the dc conductivity; we therefore also need to choose $\omega_{\max}$ in the extrapolation of the optical conductivity.
To test the
robustness of the extrapolated dc resistivity against the choice of the fitting frequency window, we also perform the quadratic fit of Eq.~\eqref{eq:quadratic_dc_extrapolation} over a wider frequency window with $\omega_{\min}=0.02t$ and $\omega_{\max} = 0.06 t$. The resulting dc resistivity differs from that obtained with $\omega_{\min}=0.025t$ and $\omega_{\max} = 0.055 t$ by at most $2\%$ for all temperatures and for both the $N=16$ and $18$ clusters.
The tests in this subsection compare different broadenings, linear and
quadratic fits, and two frequency windows. They show how sensitive the
extrapolated results are to these choices, but they do not provide a separate
estimate or an upper bound for the residual contribution of the broadened
Drude-like tail within the fitting window.

\subsection{\texorpdfstring{Two approximately linear-in-$T$ resistivity regimes}{Two approximately linear-in-T resistivity regimes}}
\label{sub:temperature_dependence_dc_resistivity}

\begin{figure}[t]
\centering
\includegraphics[width=0.8\columnwidth, keepaspectratio]{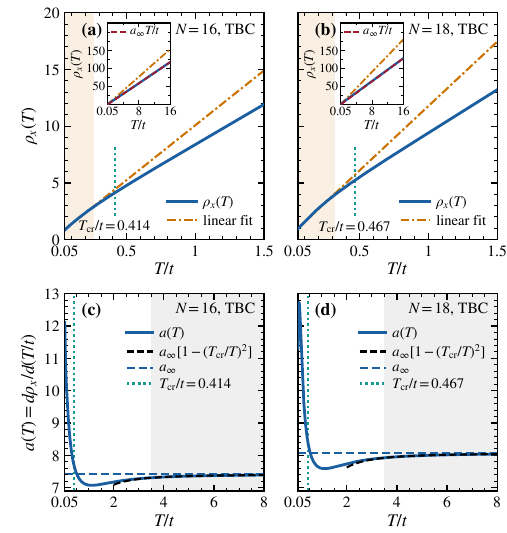}
\caption{Reproduction of Fig.~2 of the main text. (a),(b) Extrapolated dc resistivity $\rho_x(T)$ under TBCs for the $N=16$ and $18$ clusters, respectively, plotted over $0.05\leq T/t\leq1.5$. Orange dash-dotted lines are linear fits over $0.05\leq T/t\leq T_{\mathrm{cr}}/t-0.15$ (beige shading). The insets extend the range to $T/t=16$ and contrast these fits with the high-temperature asymptote $a_\infty T/t$ (red dashed lines). (c),(d) Local slope $a(T)=d\rho_x/d(T/t)$. Blue dashed lines indicate the asymptotic slope $a_{\infty}$ obtained from the ED data, and black dashed lines are fits of the form $a(T)=a_\infty[1-(T_{\mathrm{cr}}/T)^2]$ [cf.\ Eq.~\eqref{eq:constant_dos_toy_high_T_expansion_a}] over $3.5\leq T/t\leq8$ (gray shading). The resulting $T_{\mathrm{cr}}$ is indicated by green dotted lines in all panels. The high- and low-temperature approximately linear-in-$T$ regimes are connected by a gradual crossover.}
\label{fig:sm_rho}
\end{figure}

Figures~\ref{fig:sm_rho}(a) and~\ref{fig:sm_rho}(b) show the extrapolated dc resistivity $\rho_x(T)$ under TBCs for the $N=16$ and $N=18$ square clusters, respectively. The main panels show the temperature range $0.05 \leq T/t \leq 1.5$, and the insets of panels (a) and (b) show the temperature range $0.05 \leq T/t \leq 16$. We note that Fig.~\ref{fig:sm_rho} reproduces Fig.~2 of the main text for
convenience. The lowest temperature, $T/t=0.05$, is set by the sharp specific-heat peak
at $T_{C_V,1}/t\approx0.04$ for both clusters (Sec.~\ref{sub:specific_heat}),
in agreement with the crossover between a paramagnet and a regime with finite
ferromagnetic correlations found in
Ref.~\cite{fratini2026strangemetaltransportcoupling}, together with the
temperature resolution $\Delta T/t=0.01$ of the charge-transport
calculations. This ferromagnetic regime is a finite-size effect
(Sec.~\ref{sub:specific_heat}), and charge transport within it is beyond the
scope of this work.

The larger resistivity of the $N=18$ cluster at a given temperature is
consistent with its smaller carrier density, $1/18$ compared with $1/16$, since
the single doped hole is the only charge carrier.
As shown in the insets of Figs.~\ref{fig:sm_rho}(a) and~\ref{fig:sm_rho}(b), the
resistivity is linear in $T$ in the high-temperature limit $T\gg t$. We note that previous charge-transport
studies~\cite{PhysRevB.102.115142,PhysRevB.104.165143,PhysRevB.95.041110,PhysRevLett.110.086401,PhysRevB.81.054512,PhysRevB.94.235115}
have reported linear-in-$T$ resistivity of the Hubbard model at high
temperatures, and this behavior follows directly from the leading $\beta$
dependence of the Kubo expression [Eq.~\eqref{eq:sigma_regular_with_eta_L}]. We next examine how the resistivity departs from its high-temperature behavior upon cooling. The leading departure is characterized by $T_{\mathrm{cr}}/t\approx0.414$ ($N=16$) and $0.467$ ($N=18$), as obtained in
Sec.~\ref{sub:energy_resolved_f_x_function}. This temperature scale $T_{\mathrm{cr}}$ can be understood from the energy-resolved current-response function
$f_{x}(\varepsilon)$, discussed in Sec.~\ref{sub:energy_resolved_f_x_function}. In addition, at low temperatures, we find a second approximately linear-in-$T$ resistivity regime, and there is a gradual crossover between the high-temperature and low-temperature approximately linear-in-$T$ resistivity regimes. To directly compare these two approximately linear-in-$T$ resistivity regimes, we perform a linear fit $\rho_x(T) = a (T/t) + b$ over $0.05 \leq T/t \leq T_{\mathrm{cr}}/t - 0.15$ [orange
dash-dotted lines in Figs.~\ref{fig:sm_rho}(a) and~\ref{fig:sm_rho}(b)], which yields
$R^{2} \approx 0.998$ for both the $N=16$ and $18$ clusters. The fitted coefficients are
$(a,b)\approx(9.659,0.425)$ and $(11.335,0.445)$ for $N=16$ and $18$, respectively.
The maximum relative residual $|\rho_x(T)-[aT/t+b]|/\rho_x(T)$ over the fitted points
is $6.96\%$ and $6.38\%$, respectively, both at $T/t=0.05$.
Because the crossover between the
two regimes is gradual, the upper bound of this linear fitting window is kept below
$T_{\mathrm{cr}}/t$. The comparison between these two approximately linear-in-$T$ resistivity regimes is shown in the insets of Figs.~\ref{fig:sm_rho}(a) and~\ref{fig:sm_rho}(b).

In panels (c) and (d) of Fig.~\ref{fig:sm_rho}, we also show the local slope of the resistivity, $a(T) = [\rho_x(T+\Delta T) - \rho_x(T)]/(\Delta T/t)$ with $\Delta T = 0.01t$, for $0.05 \leq T/t \leq 8$. The blue dashed lines in Figs.~\ref{fig:sm_rho}(c) and~\ref{fig:sm_rho}(d) show $a_{\infty} = a(T \to \infty)$ obtained numerically for $N=16$ and $18$, respectively. We note that $a_{\infty}$ differs between $N=16$ and $18$. The local slope $a(T)$ approaches $a_{\infty}$ as the temperature increases; at
$T/t\approx 7$, for instance, $a(T)$ differs from $a_\infty$ [blue dashed lines
in Figs.~\ref{fig:sm_rho}(c) and~\ref{fig:sm_rho}(d)] by about $0.4\%$ for
$N=16$ and $0.5\%$ for $N=18$. Nevertheless, in the high-temperature regime, $a(T)$ still depends visibly on temperature. This is distinct from strictly linear-in-$T$ resistivity with a temperature-independent slope (see Sec.~\ref{sec:definition_of_f_function}), and we therefore call this regime an approximately linear-in-$T$ resistivity regime. In the second approximately linear-in-$T$ resistivity regime, the local slope $a(T)$ shows a stronger $T$ dependence; $a(T)$ is a more sensitive diagnostic than a linear fit over a chosen temperature window. In
Sec.~\ref{sec:origin_of_crossover_temperature}, we discuss the spectral origin of the crossover temperature scale $T_{\mathrm{cr}}$.

\section{Spectral origin of the crossover scale}
\label{sec:origin_of_crossover_temperature}

In Sec.~\ref{sec:numerical_results}, we found that the dc resistivity is
approximately linear in $T$ over a broad temperature range, with a gradual
crossover between a low-temperature regime and a high-temperature regime. In this
section, we discuss the spectral origin of the crossover
temperature $T_{\mathrm{cr}}$. In
Sec.~\ref{sub:energy_resolved_f_x_function}, we present numerical results for the energy-resolved
current-response function $f_x(\varepsilon)$. We identify the scale $\varepsilon_{f_x}$
over which $f_x(\varepsilon)$ departs from its approximately flat central plateau near the
spectral edges, and relate $\varepsilon_{f_x}$ to $T_{\mathrm{cr}}$ within a
simplified model. We also extract $T_{\mathrm{cr}}$ from the ED resistivity and compare it with the prediction of this simplified model. In Sec.~\ref{sub:density_of_states_and_specific_heat}, we
present numerical results for the many-body density of states (DOS). In
Sec.~\ref{sub:specific_heat}, we present numerical results for the specific heat and compare the position of its broad peak with $T_{\mathrm{cr}}$. In
Sec.~\ref{sub:comparison_between_different_energy_scales}, we discuss the connection of the present work to the related work of
Ref.~\cite{relatedwork}.

\subsection{\texorpdfstring{Energy-resolved current-response function $f_x(\varepsilon)$ and the crossover temperature $T_{\mathrm{cr}}$}{Energy-resolved current-response function fx and the crossover temperature Tcr}}
\label{sub:energy_resolved_f_x_function}

In Theorem~\ref{prop:fx_linear_T} of Sec.~\ref{sec:definition_of_f_function}, we have proved that, for any given Hamiltonian of a finite-size system, a constant dc energy-resolved current-response function $f_x(\varepsilon)$ is a necessary and sufficient condition for strictly linear-in-$T$ resistivity. Motivated by the approximately linear-in-$T$ resistivity observed for
$T/t \geq 0.05$ and by the gradual crossover between the two approximately
linear-in-$T$ regimes (Sec.~\ref{sub:temperature_dependence_dc_resistivity}),
we compute $f_x(\varepsilon)$ for the
infinite-$U$ Hubbard model with a single doped hole and examine how its deviation from a constant
is reflected in the departure of the resistivity from its high-temperature
linear form.

\begin{figure}[t]
\centering
\includegraphics[width=5.375in]{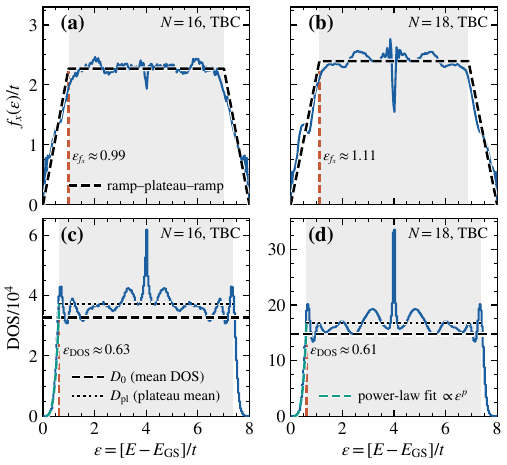}
\caption{(a),(b) Extrapolated dc $f_x(\varepsilon)$ of the $N=16$ and $18$
clusters, respectively, computed under TBCs with Lorentzian broadening
$\eta/t=0.005$ and plotted versus the excitation energy
$\varepsilon=(E-E_{\mathrm{GS}})/t$. (c),(d) Many-body density of
states (DOS). Black dashed lines indicate the two approximations of the
simplified model, namely the ramp--plateau--ramp form of
Eq.~\eqref{eq:toy_fx} in (a),(b) and a constant
DOS $D_0$, equal to the mean DOS over the full spectrum, in (c),(d). Black
dotted lines in (c),(d) mark the DOS plateau mean $D_{\mathrm{pl}}$ over
$\varepsilon_{\mathrm{DOS}}\leq\varepsilon\leq W-\varepsilon_{\mathrm{DOS}}$
(Sec.~\ref{sub:density_of_states_and_specific_heat}). Teal dashed lines in
(c),(d) show the power-law fits $D(\varepsilon)/10^4=c\varepsilon^p$ over
$ \varepsilon\leq\varepsilon_{\mathrm{DOS}}$, with $p\approx3.91$ and
$4.42$, respectively, which reach $D_{\mathrm{pl}}$ at
$\varepsilon=\varepsilon_{\mathrm{DOS}}$. Red dashed lines mark the
edge scales $\varepsilon_{f_x}\approx0.99$ and $1.11$ in (a),(b) and
$\varepsilon_{\mathrm{DOS}}\approx0.63$ and $0.61$ in (c),(d), respectively.
Gray shaded regions indicate the corresponding plateau regions,
$[\varepsilon_{f_x},W-\varepsilon_{f_x}]$ in (a),(b) and
$[\varepsilon_{\mathrm{DOS}},W-\varepsilon_{\mathrm{DOS}}]$ in (c),(d).}
\label{fig:fx_dos_N16_N18_TBC}
\end{figure}

Recall that $\varepsilon=(E-E_{\mathrm{GS}})/t$ with $E_{\mathrm{GS}}=-4t$, and
$\widetilde{T}\equiv T/t$. In the numerical calculations of the energy-resolved current-response function $f_x$ under TBCs, we first divide the energy spectrum $0\leq \varepsilon \leq W$ with $W=8$ into the following energy bins $b_i$. Within $0.04$ of either spectral edge, i.e., over $[0,0.04)$ and $[7.96,8]$, we use the bin width $\Delta \varepsilon = 0.01$, i.e., the energy bins $b_i = [0,0.01), [0.01,0.02), [0.02,0.03), [0.03,0.04)$ and $[7.96, 7.97), [7.97, 7.98), [7.98, 7.99), [7.99, 8]$ with bin centers $\varepsilon^c_i = 0.005,0.015,0.025,0.035, 7.965,7.975, 7.985, 7.995$. Over $[0.04, 7.96)$, we use the bin width $\Delta \varepsilon = 0.04$, i.e., $b_i = [0.04,0.08), [0.08,0.12), \ldots, [7.92, 7.96)$ with bin centers $\varepsilon^c_i = 0.06, 0.10, \ldots, 7.94$. For each twist and a given energy bin $b_i$ with center $\varepsilon^c_i$, we first evaluate $f_x(\ket{n},\omega,\eta)$ of Eq.~\eqref{eq:fx_state_resolved} with broadening $\eta = 0.005t$ for all eigenstates $\ket{n}$ whose eigenenergies lie in $b_i$. We then average these $f_x(\ket{n},\omega,\eta)$ within $b_i$ to obtain $f_{x, \boldsymbol{\theta}}(\varepsilon^c_i, \omega, \eta)$, and average over all $N_\theta = 121$ twists: $f_{x}(\varepsilon^c_i, \omega, \eta) = N_{\theta}^{-1} \sum_{\boldsymbol{\theta}} f_{x, \boldsymbol{\theta}}(\varepsilon^c_i, \omega, \eta)$. Finally, we fit $f_{x}(\varepsilon^c_i,\omega,\eta)$ at the seven frequencies
$\omega/t=0.025,0.030,\ldots,0.055$ to the quadratic form
$f_{x}(\varepsilon^c_i,\omega,\eta)=f_x(\varepsilon^c_i)+c_2(\varepsilon^c_i)\,(\omega/t)^2$,
and take the intercept $f_x(\varepsilon^c_i)$ as the dc value. The resulting $f_x(\varepsilon)$ for $N=16$ and $18$ is shown in
Figs.~\ref{fig:fx_dos_N16_N18_TBC}(a) and~\ref{fig:fx_dos_N16_N18_TBC}(b), respectively.

For both the $N=16$ and $18$ clusters, $f_x(\varepsilon)$ forms a broad,
approximately flat central plateau over
$\varepsilon_{f_x}\leq\varepsilon\leq W-\varepsilon_{f_x}$, with the edge
scale $\varepsilon_{f_x}\approx0.99$ and $1.11$ for $N=16$ and $18$,
respectively [red dashed lines in Figs.~\ref{fig:fx_dos_N16_N18_TBC}(a)
and~\ref{fig:fx_dos_N16_N18_TBC}(b)]. Near the spectral edges,
$f_x(\varepsilon)$ varies more strongly with energy. We determine $f_0$ and $\varepsilon_{f_x}$ by requiring that $f_0$ be the mean of
$f_x(\varepsilon)$ over $\varepsilon_{f_x}\leq\varepsilon\leq W-\varepsilon_{f_x}$ and that the
ramp--plateau--ramp profile of Eq.~\eqref{eq:toy_fx} have the same area as
$f_x(\varepsilon)$, i.e., $f_0(W-\varepsilon_{f_x})=\int_0^W d\varepsilon\,f_x(\varepsilon)$.
This gives $f_0\approx2.270$ and $2.397$ for $N=16$ and $18$, respectively. We note that some local
structure remains within the plateau, for example the dip of
$f_x(\varepsilon)$ near the band center for $N=18$ in
Fig.~\ref{fig:fx_dos_N16_N18_TBC}(b). We do not yet have an analytical
understanding of this dip. Whether the edge scale $\varepsilon_{f_x}$ is
related to the eigenstate thermalization hypothesis also remains an open
question.

\begin{figure}[t]
\centering
\includegraphics[width=4.375in]{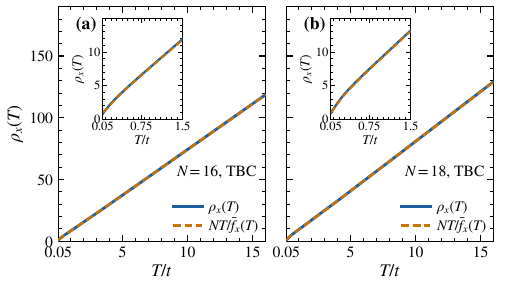}
\caption{Comparison between the dc resistivity and its reconstruction from the dc
$f_x(\varepsilon)$ and the DOS [Eq.~\eqref{eq:sm_model_dcconductivty}].
(a) $N=16$ and (b) $N=18$ clusters under TBCs, for $0.05\leq T/t\leq16$. The
insets enlarge the range $0.05\leq T/t\leq1.5$. Blue solid lines: dc
resistivity $\rho_x(T)$ of Fig.~\ref{fig:sm_rho}, obtained from the quadratic
extrapolation of the twist-averaged optical conductivity
[Eq.~\eqref{eq:quadratic_dc_extrapolation}]. Orange dashed lines:
$NT/\overline{f}_x(T)$ of Eq.~\eqref{eq:sm_model_dcconductivty}, where
$\overline{f}_x(T)$ is the thermal average of the dc
$f_x(\varepsilon)$ over the twist-averaged DOS of
Figs.~\ref{fig:fx_dos_N16_N18_TBC}(c) and~\ref{fig:fx_dos_N16_N18_TBC}(d).
The dc $f_x(\varepsilon)$ is obtained by fitting
$f_x(\varepsilon,\omega,\eta=0.005t)$ at the seven frequencies
$\omega/t=0.025,0.030,\ldots,0.055$ to
$f_x(\varepsilon)+c_2(\varepsilon)(\omega/t)^2$. The blue and orange curves
have the same high-temperature slope. They differ by less than $1\%$ for
$T/t\gtrsim1$ and by at most $2.5\%$ ($N=16$) and $5.9\%$ ($N=18$) for
$0.1\leq T/t\leq1$. At the lowest temperature, $T/t=0.05$, the difference is
$32\%$ and $16\%$, respectively.}
\label{fig:rho_fx_dos_comparison}
\end{figure}

Moreover, to clarify the connection between $\varepsilon_{f_x}$ and the
departure of the resistivity from its high-temperature behavior, we consider
the approximate dc resistivity
\begin{align}
\rho_x(T)=\frac{NT}{\overline{f}_x(T)},\qquad
\overline{f}_x(T)
=\frac{\displaystyle\int_0^W d\varepsilon\,
D(\varepsilon)f_x(\varepsilon)e^{-\varepsilon/\widetilde{T}}}
{\displaystyle\int_0^W d\varepsilon\,
D(\varepsilon)e^{-\varepsilon/\widetilde{T}}},
\label{eq:sm_model_dcconductivty}
\end{align}
where we have used $\lim_{\omega\to0}(1-e^{-\beta\omega})/\omega=\beta$, $\widetilde{T}\equiv T/t$,
$f_x(\varepsilon)$ is the twist-averaged $f_x$ extrapolated to $\omega\to0$,
and $D(\varepsilon)$ is the twist-averaged density of states (DOS) of
Sec.~\ref{sub:density_of_states_and_specific_heat}. We note that
Eq.~\eqref{eq:sm_model_dcconductivty} does not, in general, reproduce the dc
resistivity obtained from the twist-averaged conductivity in
Sec.~\ref{sub:temperature_dependence_dc_resistivity}, because the latter is
obtained by evaluating the finite-frequency conductivity for each twist from
$f_{x,\boldsymbol{\theta}}(\varepsilon,\omega,\eta)$ with its own DOS and partition
function $Z_{\boldsymbol{\theta}}(T)$, averaging over twists,
and then extracting the dc conductivity by the quadratic
finite-frequency fit before taking its inverse. Nevertheless, as shown in
Fig.~\ref{fig:rho_fx_dos_comparison}, for the infinite-$U$ Hubbard model with
a single doped hole, the resistivity given by
Eq.~\eqref{eq:sm_model_dcconductivty} is close to that obtained from the
twist-averaged conductivity for $T/t\gtrsim0.1$. The relative differences
between the two in different temperature ranges are listed in the caption of
Fig.~\ref{fig:rho_fx_dos_comparison}.

Furthermore, we consider an approximate model with two further approximations: (1) a constant DOS over the entire spectrum $0 \leq \varepsilon \leq W$ and (2) a symmetric ramp--plateau--ramp form of $f_x(\varepsilon)$,
\begin{equation}
f_x(\varepsilon)=
\begin{cases}
f_0\,\varepsilon/\varepsilon_{f_x}, & 0 \leq \varepsilon<\varepsilon_{f_x},\\
f_0, & \varepsilon_{f_x}\leq \varepsilon \leq W-\varepsilon_{f_x},\\
f_0\,(W-\varepsilon)/\varepsilon_{f_x}, & W-\varepsilon_{f_x}<\varepsilon\leq W,
\end{cases}
\label{eq:toy_fx}
\end{equation}
with plateau value $f_0$. See Fig.~\ref{fig:fx_dos_N16_N18_TBC} for a schematic of these approximations. We adopt the constant-DOS approximation for two reasons. First, in the proof in Sec.~\ref{sec:definition_of_f_function}, we have shown that, regardless of how the degeneracy, or equivalently the DOS, depends on $\varepsilon$, a constant $f_x(\varepsilon)$ is a necessary and sufficient condition for strictly linear-in-$T$ resistivity. It is therefore natural to consider a simplified model that isolates the effect of $\varepsilon_{f_x}$. Second, as shown in Figs.~\ref{fig:fx_dos_N16_N18_TBC}(c) and~\ref{fig:fx_dos_N16_N18_TBC}(d)
and discussed in Sec.~\ref{sub:density_of_states_and_specific_heat}, the many-body DOS of both the $N=16$ and $18$ clusters is also approximately flat over a broad central region of the many-body spectrum, $\varepsilon_{\mathrm{DOS}} \leq \varepsilon \leq W- \varepsilon_{\mathrm{DOS}}$, with $\varepsilon_{\mathrm{DOS}} < \varepsilon_{f_x}$ for each $N$. The effects of these two approximations are discussed after Eq.~\eqref{eq:constant_dos_curvature_scale}.

With these three approximations (Eq.~\eqref{eq:sm_model_dcconductivty}, Eq.~\eqref{eq:toy_fx}, and constant DOS), we now discuss how
$\varepsilon_{f_x}$ controls the departure of the resistivity
from its high-temperature behavior. The thermally averaged $f_x(\varepsilon)$ [Eq.~\eqref{eq:sm_model_dcconductivty}] is
the continuum version of the level sums in
Eqs.~\eqref{eq:sigma_dc_level_resolved} and~\eqref{eq:rho_level_resolved}. Under the constant-DOS approximation, $D(\varepsilon)=D_0$ cancels between the numerator and
denominator in $\overline{f}_x(T)$ [Eq.~\eqref{eq:sm_model_dcconductivty}], and the two integrals in
$\overline{f}_x(T)$ [Eq.~\eqref{eq:sm_model_dcconductivty}] become
\begin{align}
\int_0^W d\varepsilon\,e^{-\varepsilon/\widetilde{T}}
&=\widetilde{T}\left(1-e^{-W/\widetilde{T}}\right),
\notag\\
\int_0^W d\varepsilon\,f_x(\varepsilon)e^{-\varepsilon/\widetilde{T}}
&=\frac{f_0\widetilde{T}^2}{\varepsilon_{f_x}}
\left(1-e^{-\varepsilon_{f_x}/\widetilde{T}}\right)
\left(1-e^{-(W-\varepsilon_{f_x})/\widetilde{T}}\right).
\label{eq:constant_dos_integrals}
\end{align}
Consequently, the dc resistivity at temperature $T$ is
\begin{equation}
\rho_x(T)=\frac{NT}{\overline{f}_x(T)}=\frac{Nt\varepsilon_{f_x}}{f_0}
\frac{1-e^{-W/\widetilde{T}}}
{\left(1-e^{-\varepsilon_{f_x}/\widetilde{T}}\right)
 \left(1-e^{-(W-\varepsilon_{f_x})/\widetilde{T}}\right)}.
\label{eq:constant_dos_toy_resistivity}
\end{equation}
In the high-temperature limit, we have 
\begin{equation}
\rho_x(T\to\infty)=a_\infty \frac{T}{t},
\qquad
a_{\infty} = \frac{NWt}{f_0 (W - \varepsilon_{f_x})},
\label{eq:ainfty}
\end{equation}
where $a_{\infty}$ is the local slope of the dc resistivity in the high-temperature limit $T/t \to \infty$ under the above approximations.
The local slope at finite temperature is
\begin{equation}
a(T)\equiv\frac{d\rho_x}{d(T/t)}
=\frac{\rho_x(T)}{\widetilde{T}^2}
\left[
\frac{\varepsilon_{f_x}}{e^{\varepsilon_{f_x}/\widetilde{T}}-1}
+\frac{W-\varepsilon_{f_x}}{e^{(W-\varepsilon_{f_x})/\widetilde{T}}-1}
-\frac{W}{e^{W/\widetilde{T}}-1}
\right].
\label{eq:constant_dos_toy_local_slope}
\end{equation}
Finally, the high-temperature expansions of
Eqs.~\eqref{eq:constant_dos_toy_resistivity}
and~\eqref{eq:constant_dos_toy_local_slope} are
\begin{align}
\rho_x(T)&=a_\infty \frac{T}{t}
\left[
1+\frac{\varepsilon_{f_x}(W-\varepsilon_{f_x})}{12\widetilde{T}^2}
+\mathcal{O}(\widetilde{T}^{-4})
\right]
= a_\infty \frac{T}{t}
\left[
1+\left(\frac{T_{\mathrm{cr}}}{T}\right)^{2}
+\mathcal{O}(\widetilde{T}^{-4})
\right],
\label{eq:constant_dos_toy_high_T_expansion}
\\
a(T) &= a_{\infty}
\left[
1 - \left(\frac{T_{\mathrm{cr}}}{T}\right)^{2} + \mathcal{O}(\widetilde{T}^{-4})
\right],
\label{eq:constant_dos_toy_high_T_expansion_a}
\end{align}
where
\begin{equation}
\frac{T_{\mathrm{cr}}}{t}
=\sqrt{\frac{\varepsilon_{f_x}(W-\varepsilon_{f_x})}{12}}.
\label{eq:constant_dos_curvature_scale}
\end{equation}
According to Eqs.~\eqref{eq:constant_dos_toy_high_T_expansion}
and~\eqref{eq:constant_dos_toy_high_T_expansion_a}, $T_{\mathrm{cr}}$ is the temperature scale that controls the leading deviation of the resistivity from its
high-temperature asymptote upon cooling. Only even powers of $1/\widetilde{T}$ appear in these expansions, because both the DOS (under the constant-DOS approximation) and
$f_x(\varepsilon)$ [under the approximation in Eq.~\eqref{eq:toy_fx}] are symmetric about the band center $\varepsilon=W/2$ with $W=8$.

Next, we compare $T_{\mathrm{cr}}$ and $a_{\infty}$ obtained from the numerical resistivity with those predicted by Eqs.~\eqref{eq:constant_dos_curvature_scale} and~\eqref{eq:ainfty}, and then discuss the effects of the ramp--plateau--ramp and constant-DOS approximations used in deriving these equations. For each of the $N=16$ and $18$ clusters, we fit the local slope of the extrapolated dc resistivity to
$a(T)=a_\infty[1-(T_{\mathrm{cr}}/T)^2]$ over $3.5\leq T/t\leq8$ [black dashed lines in Figs.~\ref{fig:sm_rho}(c) and~\ref{fig:sm_rho}(d)], treating
$a_\infty$ and $T_{\mathrm{cr}}$ as fitting parameters. This gives
$T_{\mathrm{cr}}/t\approx0.414$ and $a_{\infty} \approx 7.412$ for $N=16$, and $T_{\mathrm{cr}}/t\approx0.467$ and $a_{\infty} \approx 8.068$ for $N=18$. Substituting $W=8$, $\varepsilon_{f_x} \approx 0.99$ and $1.11$, and $f_0 \approx 2.270$ and $2.397$ for $N=16$ and $18$, respectively, into Eqs.~\eqref{eq:constant_dos_curvature_scale} and~\eqref{eq:ainfty}, we obtain $T_{\mathrm{cr}}^{\mathrm{th}}/t \approx 0.760$ and $a_{\infty}^{\mathrm{th}} \approx 8.042$ for $N=16$, and $T_{\mathrm{cr}}^{\mathrm{th}}/t \approx 0.799$ and $a_{\infty}^{\mathrm{th}} \approx 8.724$ for $N=18$. Both $T_{\mathrm{cr}}$ and $a_{\infty}$ obtained from the numerical results are therefore smaller than those obtained with the two approximations. The direction of both differences can be understood as follows. Because $\varepsilon_{\mathrm{DOS}} < \varepsilon_{f_x}$, the constant-DOS approximation assigns a larger weight to the edge regions $\varepsilon<\varepsilon_{\mathrm{DOS}}$ and $\varepsilon>W-\varepsilon_{\mathrm{DOS}}$, where $f_x(\varepsilon)$ drops rapidly. This makes $\overline{f}_x(T)$ in Eq.~\eqref{eq:sm_model_dcconductivty} smaller than its actual value, and the resistivity in Eq.~\eqref{eq:constant_dos_toy_resistivity}, including $a_{\infty}$, larger than the actual one. The scale $T_{\mathrm{cr}}$ arises from the deviation of $f_{x}(\varepsilon)$ from a
constant in the edge regions of width $\varepsilon_{f_x}$. Since
$\varepsilon_{\mathrm{DOS}}<\varepsilon_{f_x}$ (Sec.~\ref{sub:density_of_states_and_specific_heat}), the actual DOS is suppressed in
part of these edge regions, which carry less thermal weight than
under a constant DOS. This tends to reduce $T_{\mathrm{cr}}$. We note that this intuitive understanding relies on the above approximations. Because of the complicated energy dependence of the actual $f_x(\varepsilon)$ and DOS, we do not have more accurate predictions for $T_{\mathrm{cr}}$ and $a_{\infty}$.

To test the dependence of $T_{\mathrm{cr}}$ on the fitting window, we also change the fitting window to
$3\leq T/t\leq8$, which gives
$T_{\mathrm{cr}}/t\approx0.405$ and $0.457$ for $N=16$ and $18$, respectively,
i.e., $2.19\%$ and $2.13\%$ below those obtained with the fitting window $3.5\leq T/t\leq8$.
Extending the window to $3.5\leq T/t\leq20$ instead gives
$T_{\mathrm{cr}}/t\approx0.425$ and $0.479$, which are $2.65\%$ and $2.58\%$
above those obtained with the fitting window $3.5\leq T/t\leq8$.

\subsection{Many-body density of states}
\label{sub:density_of_states_and_specific_heat}

Figures~\ref{fig:fx_dos_N16_N18_TBC}(c) and~\ref{fig:fx_dos_N16_N18_TBC}(d)
show the many-body DOS $D(\varepsilon)$ for $N=16$ and $18$ under TBCs. For
the single-hole case considered here, the full Hilbert
space dimension, i.e., the total number of eigenstates, is $N2^{N-1}$.
As mentioned at the beginning of Sec.~\ref{sec:numerical_results}, we retain
only the positive-$S^z$ sectors, reducing
the number of states included in the calculation from $N2^{N-1}$ to $N2^{N-2}$, i.e.,
$262144$ and $1179648$ states for $N=16$ and $18$, respectively. To present the numerical DOS, we use the
same energy bins as introduced in
Sec.~\ref{sub:energy_resolved_f_x_function}. For each bin $b_i$, the DOS is
obtained by dividing the twist-averaged state count by the bin width,
$D_i=(N_{\theta}\Delta\varepsilon_i)^{-1}
\sum_{\boldsymbol{\theta}} n_{\boldsymbol{\theta},i}$,
where $n_{\boldsymbol{\theta},i}$ is the number of retained
eigenstates in bin $b_i$ at twist $\boldsymbol{\theta}$.

From Figs.~\ref{fig:fx_dos_N16_N18_TBC}(c) and~\ref{fig:fx_dos_N16_N18_TBC}(d)
we make three numerical observations. (1) The DOS has a peak at the band
center $\varepsilon=W/2$ (i.e., $E=0$) for both $N=16$ and $18$. (2) The DOS
is approximately flat for
$\varepsilon_{\mathrm{DOS}}\leq\varepsilon\leq W-\varepsilon_{\mathrm{DOS}}$,
with the edge scale $\varepsilon_{\mathrm{DOS}}\approx0.629$ for $N=16$ and
$0.610$ for $N=18$. (3) For $\varepsilon<\varepsilon_{\mathrm{DOS}}$, the DOS
is well described by the power law $D(\varepsilon)/10^4=c\,\varepsilon^p$,
with $c=22.84\pm4.62$ and $p=3.91\pm0.16$ for $N=16$, and
$c=148.61\pm33.10$ and $p=4.42\pm0.17$ for $N=18$. The parameters $c$, $p$,
and $\varepsilon_{\mathrm{DOS}}$ are determined as follows. We obtain $c$ and
$p$ from an equal-weight least-squares fit of $\ln(D_i/10^4)$ to
$\ln c+p\ln\varepsilon_i$, using the bins with $D_i>0$ in the range
$0.10\leq\varepsilon_i\leq\varepsilon_{\mathrm{DOS}}$. The edge scale
$\varepsilon_{\mathrm{DOS}}$ is determined by the condition
$10^4c\,\varepsilon_{\mathrm{DOS}}^{p}=D_{\mathrm{pl}}$, where the plateau
value $D_{\mathrm{pl}}$ is the bin-width-weighted mean of the DOS over the
bins whose centers lie in
$[\varepsilon_{\mathrm{DOS}},W-\varepsilon_{\mathrm{DOS}}]$. Since both the
fit range and the plateau window depend on $\varepsilon_{\mathrm{DOS}}$, the
fit and the plateau mean are iterated to convergence. The quoted uncertainty
of $p$ is its regression standard error, and that of $c$ is the propagated
error $s_c=c\,s_{\ln c}$, where $s_{\ln c}$ is the standard error of the
fitted intercept $\ln c$.
We use these fitted power laws to model the DOS edges in the specific-heat
analysis of Sec.~\ref{sub:specific_heat} [Eq.~\eqref{eq:dos_model}]. In the
simplified transport model of Sec.~\ref{sub:energy_resolved_f_x_function}
[Eqs.~\eqref{eq:toy_fx}--\eqref{eq:constant_dos_curvature_scale}], we instead
approximate the DOS by a constant to isolate the role of $\varepsilon_{f_x}$.

We note that the DOS computed here is a many-body DOS, so the numerically obtained exponent $p$ cannot
be inferred from a single-particle dispersion. For comparison, we recall the
standard single-particle relation between the DOS and the dispersion. Consider an isotropic low-energy
single-particle dispersion in $d$ spatial dimensions,
\begin{equation}
\Delta E(\mathbf{k})
\equiv
E(\mathbf{k})-E(\mathbf{k}_0)
=
A\lvert \mathbf{k}-\mathbf{k}_0\rvert^z,
\label{eq:low_energy_dispersion}
\end{equation}
where $\mathbf{k}_0$ denotes the location of the band minimum, $A$ is a
constant, and $z$ is the dynamical exponent. The corresponding low-energy single-particle DOS scales as
$D_{\mathrm{sp}}(\Delta E)\propto(\Delta E)^{p_{\mathrm{sp}}}$ with $p_{\mathrm{sp}}=d/z-1$, which for $d=2$ reduces to $p_{\mathrm{sp}}=2/z-1$. For the many-body DOS of the infinite-$U$ Hubbard model with a single doped
hole, it remains an open question whether the exponent $p$ and the origin of
the peak at the band center can be understood analytically, either for
finite-size systems or in the thermodynamic limit.

\subsection{Specific heat}
\label{sub:specific_heat}

\begin{figure}[t]
\centering
\includegraphics[width=5.175in]{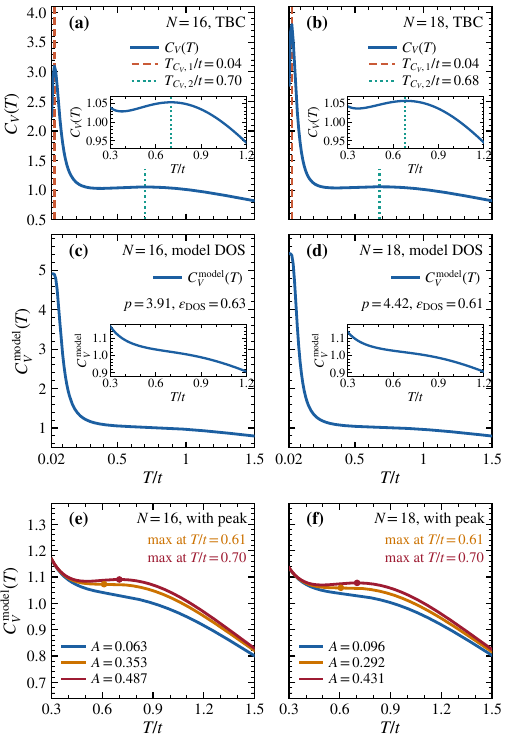}
\caption{(a),(b) Reproduction of Fig.~3 of the main text: twist-averaged specific
heat $C_V(T)$ [Eq.~\eqref{eq:CV_TBCs}] for $N=16$ and $18$, respectively. Dashed
and dotted lines mark the sharp low-temperature peak at $T_{C_V,1}$ and the broad
peak at $T_{C_V,2}$, respectively. (c),(d) Specific heat $C_V^{\mathrm{model}}(T)$
of the simplified DOS of Eq.~\eqref{eq:dos_model}, with $p=3.91$ and
$\varepsilon_{\mathrm{DOS}}=0.63$ for $N=16$ and $p=4.42$ and
$\varepsilon_{\mathrm{DOS}}=0.61$ for $N=18$; it decreases monotonically and shows
no peak. The insets in (a)--(d) enlarge the interval $0.3\leq T/t\leq1.2$.
(e),(f) Specific heat $C_V^{\mathrm{model}}(T)$ of the DOS with a band-center peak
[Eq.~\eqref{eq:dos_model_peak}, $\sigma=0.03$] for three peak weights $A$, shown over
$0.3\leq T/t\leq1.5$. The blue curves use the weights extracted from the numerical DOS
in Figs.~\ref{fig:fx_dos_N16_N18_TBC}(c) and~\ref{fig:fx_dos_N16_N18_TBC}(d), $A=0.063$ ($N=16$) and $0.096$ ($N=18$), i.e.,
about $0.9\%$ and $1.4\%$ of all states; $C_V^{\mathrm{model}}(T)$ remains monotonic, i.e., no peak. If we further increase $A$ in Eq.~\eqref{eq:dos_model_peak}, the specific heat develops a broad peak, as shown by the orange and red curves in (e) and (f), where the dots mark the peak positions.
}
\label{fig:comparison}
\end{figure}

We also compute the total specific heat for both the $N=16$ and $18$ clusters under TBCs. For each twist $\boldsymbol{\theta}$, we first evaluate $C_V^{(\boldsymbol{\theta})}(T)$ within the canonical single-hole sector,
\begin{align}
C_V^{(\boldsymbol{\theta})}(T)
=
\beta^2\left[
\langle E^2\rangle_{\boldsymbol{\theta}}
-
\langle E\rangle_{\boldsymbol{\theta}}^2
\right],
\end{align}
where $\langle\cdots\rangle_{\boldsymbol{\theta}}$ denotes the thermal average at twist $\boldsymbol{\theta}$. We then average over the $N_{\theta}=121$ twists,
\begin{align}
C_V(T)
=
\frac{1}{N_{\theta}}
\sum_{\boldsymbol{\theta}}
C_V^{(\boldsymbol{\theta})}(T).
\label{eq:CV_TBCs}
\end{align}
The twist-averaged specific heat $C_V(T)$ [Eq.~\eqref{eq:CV_TBCs}] is plotted
in Figs.~\ref{fig:comparison}(a) and~\ref{fig:comparison}(b) for $N=16$ and
$18$, respectively.
The specific heat $C_{V}(T)$ exhibits two peaks for both $N=16$ and $18$. The first peak is
sharp and lies at a very low temperature, $T_{C_V,1}/t\approx0.04$, for both
clusters. Previous work~\cite{fratini2026strangemetaltransportcoupling}
reports that, upon cooling, the system crosses over from a paramagnet to a ferromagnetic
regime, and the sharp
peak at $T_{C_V,1}$ is consistent with this magnetic crossover. This crossover is a finite-size effect, since the infinite-$U$ Hubbard model
with a single doped hole is paramagnetic at any $T>0$ in the thermodynamic
limit~\cite{PhysRevB.43.8779}, and $T_{C_V,1}$ is therefore expected to
vanish as $N\to\infty$. This peak also sets the lowest temperature,
$T/t=0.05$, of our transport calculations
(Sec.~\ref{sub:temperature_dependence_dc_resistivity}).
The second peak is much broader and
is located at $T_{C_V,2}/t\approx0.70$ ($N=16$) and $0.68$ ($N=18$); the insets of
Figs.~\ref{fig:comparison}(a) and~\ref{fig:comparison}(b) enlarge the interval $0.3 \leq T/t \leq 1.2$ to make this broad peak clear. The position of this second peak is comparable to $T_{\mathrm{cr}}$ obtained in Sec.~\ref{sub:energy_resolved_f_x_function}, although the two are not the same. This second peak of the specific heat thus also supports a gradual crossover between the high-temperature and low-temperature approximately linear-in-$T$ resistivity regimes discussed in Sec.~\ref{sub:temperature_dependence_dc_resistivity}.

Since the specific heat at each twist is determined by the corresponding DOS, and since $\varepsilon_{\mathrm{DOS}}$ ($\approx0.63$ and $0.61$; see Sec.~\ref{sub:density_of_states_and_specific_heat}) is numerically
close to $T_{C_V,2}/t$ ($\approx0.70$ and $0.68$), we ask whether $\varepsilon_{\mathrm{DOS}}$ and $T_{C_V,2}/t$ are
connected. To examine this, we consider a simplified DOS based on
the discussion in
Sec.~\ref{sub:density_of_states_and_specific_heat},
\begin{equation}
D_{\mathrm{model}}(\varepsilon)=
\begin{cases}
D_0(\varepsilon/\varepsilon_{\mathrm{DOS}})^p, & 0\leq\varepsilon<\varepsilon_{\mathrm{DOS}},\\
D_0, & \varepsilon_{\mathrm{DOS}}\leq\varepsilon\leq W-\varepsilon_{\mathrm{DOS}},\\
D_0[(W-\varepsilon)/\varepsilon_{\mathrm{DOS}}]^p, & W-\varepsilon_{\mathrm{DOS}}<\varepsilon\leq W,
\end{cases}
\label{eq:dos_model}
\end{equation}
where $D_0$ is an overall normalization factor that cancels in the thermal
averages. Using the fitted values of $p$ and $\varepsilon_{\mathrm{DOS}}$ (Sec.~\ref{sub:density_of_states_and_specific_heat}) for
$N=16$ and $18$, we evaluate $C_V(T)$ over $0.05\leq T/t\leq1.5$ (see Figs.~\ref{fig:comparison}(c) and~\ref{fig:comparison}(d)) and find
no peak in $C_V(T)$ corresponding to the broad second peak observed in the ED
results.

We next include the DOS peak at the band center, $\varepsilon=W/2$ (i.e.,
$E=0$), observed in Figs.~\ref{fig:fx_dos_N16_N18_TBC}(c)
and~\ref{fig:fx_dos_N16_N18_TBC}(d), by adding a narrow Gaussian peak of
weight $A$ to Eq.~\eqref{eq:dos_model},
\begin{equation}
D_{\mathrm{model}}^{\prime}(\varepsilon)=D_{\mathrm{model}}(\varepsilon)
+A\,D_0\,g_\sigma\!\left(\varepsilon-W/2\right),\qquad
g_\sigma(x)=\frac{1}{\sqrt{2\pi\sigma^2}}\,e^{-x^2/(2\sigma^2)},
\label{eq:dos_model_peak}
\end{equation}
so that the peak contains $AD_0$ states. The overall factor $D_0$ still
cancels in the thermal averages. Because the DOS peak at the band center is
narrower than the bin width, we take $\sigma=0.03$. Identifying $D_0$ with
the plateau mean $D_{\mathrm{pl}}$ of the numerical DOS, we determine $A$
from the excess of the numerical DOS over $D_{\mathrm{pl}}$ in the four bins
with $|\varepsilon-W/2|\leq0.08$. This gives $A\approx0.063$ for $N=16$ and
$0.096$ for $N=18$, for which the added peak contains about $0.9\%$ and
$1.4\%$ of all states. For these values of $A$, however, $C_V(T)$ obtained
from Eq.~\eqref{eq:dos_model_peak} remains monotonic and shows no broad peak
[blue curves in Figs.~\ref{fig:comparison}(e) and~\ref{fig:comparison}(f)].
If we further increase $A$ in Eq.~\eqref{eq:dos_model_peak}, a broad peak of
the specific heat appears [Figs.~\ref{fig:comparison}(e)
and~\ref{fig:comparison}(f)] once the added peak contains more than about
$4.8\%$ ($N=16$) and $4.0\%$ ($N=18$) of all states, i.e., for values of $A$
about $5.5$ and $3$ times those obtained from the numerical DOS. The peak
weight required to produce this broad peak depends on the simplifications of
the model DOS, namely the power-law edges and the constant plateau, and may
differ for the actual DOS. The central peak of $D(\varepsilon)$ is therefore
a possible origin of the second, broad peak of the specific heat at
$T_{C_V,2}$, although the weight extracted from the numerical DOS is not
sufficient within the simplified model of Eq.~\eqref{eq:dos_model_peak}. We
note that we do not yet have an analytical understanding of why the
many-body DOS exhibits a peak at the band center.

Finally, we note that the infinite-$U$ Hubbard model with a single doped hole should not be directly connected to the strange-metal phenomena observed at a finite density of charge carriers. For completeness, we nevertheless recall the specific-heat behavior typically expected at a finite carrier density. In a conventional Landau Fermi liquid, the low-temperature specific heat satisfies $C_V(T)=\gamma T+o(T)$, so that $C_{V}(T)/T$ approaches a finite constant~\cite{PhysRevLett.95.026402}.
In certain quantum-critical strange metals, the specific-heat coefficient instead exhibits a logarithmic enhancement, $C_V(T)/T=\gamma_0+A'\ln(T_0/T)$ with $A'>0$, as observed in quantum-critical heavy-fermion and cuprate systems~\cite{PhysRevLett.72.3262, michonThermodynamicSignaturesQuantum2019}.
Our single-hole results on finite clusters do not determine the specific heat
of the infinite-$U$ Hubbard model at a finite hole density in the
thermodynamic limit.

\subsection{Connection to a related work}
\label{sub:comparison_between_different_energy_scales}

While completing this manuscript, we became aware of Ref.~\cite{relatedwork}, which also investigates charge transport in the infinite-$U$ Hubbard
model on the square lattice with a single doped
hole. The authors of Ref.~\cite{relatedwork} also identify two regimes, which they call the
semi-quantum bad metal (corresponding to the low-temperature approximately linear-in-$T$ resistivity regime identified in this work) and the high-temperature
bad metal (corresponding to the high-temperature approximately linear-in-$T$ resistivity regime identified in this work). Their identification of the two approximately linear-in-$T$ resistivity regimes is based on the Nernst--Einstein relation~\cite{relatedwork},
\begin{align}
\rho_x(T) = \frac{1}{\chi_c D_c},
\label{eq:nernst_einstein}
\end{align}
where $\chi_c$ is the charge compressibility, whose inverse is estimated in a finite-size system
by the finite difference
\begin{equation}
\chi_c^{-1}(T)=
N\left[F_{N_e+1}(T)-2F_{N_e}(T)+F_{N_e-1}(T)\right].
\label{eq:charge_compressibility}
\end{equation}
Here, $F_{N_e}(T)$ denotes the canonical free energy in the
sector with $N_e$ electrons, averaged over the chosen
twists. We note that Ref.~\cite{relatedwork} uses twists different from those in this work. For the single-hole sector, $N_e=N-1$, so the difference involves the
zero-, one-, and two-hole sectors, i.e., $F_{16}$, $F_{15}$, and $F_{14}$ for
$N=16$, and $F_{18}$, $F_{17}$, and $F_{16}$ for $N=18$. The charge
diffusivity is then inferred from Eq.~\eqref{eq:nernst_einstein} as
$D_c=1/(\rho_x\chi_c)$.

Reference~\cite{relatedwork} finds that, although the resistivity is approximately linear in $T$ in
both regimes, the compressibility $\chi_c$ scales as $1/T$ in the high-temperature bad-metal regime and approaches a nearly constant value in the semi-quantum bad-metal regime. Correspondingly, the
charge diffusivity $D_c$ is approximately constant in the high-temperature bad-metal regime, while
$D_c^{-1}$ becomes approximately linear in $T$ in the semi-quantum bad-metal regime.
Operationally, the authors of Ref.~\cite{relatedwork} identify the crossover temperature
$T_{\mathrm{quant}}/t\approx0.5$--$0.8$ from departures from high-temperature
behavior, including the saturation of $1/\chi_c$ toward a nearly constant
value. They also compute the specific heat and find two peaks, with the broad
higher-temperature peak at
$T_{C_V,2}/t\approx0.75$--$0.8$; the slight difference from our $T_{C_V,2}/t$ may arise from the different twists used in the two calculations. Our work provides a complementary perspective on the crossover between the two approximately linear-in-$T$ resistivity regimes:
Ref.~\cite{relatedwork} identifies the two regimes through the different temperature
dependences of the compressibility, whereas we trace the crossover to the energy-resolved current-response function $f_x(\varepsilon)$.

\end{document}